# Analysis and prediction of changes in the temperature of the pure freshwater ice column in the Antarctic and the Arctic

A.A. Fedotov, V.V. Kaniber, P.V. Khrapov

**Abstract** – This paper investigates the initial boundary value problem for a non-stationary one-dimensional heat equation that simulates the temperature distribution in freshwater ice near the Earth's poles. The mathematical model has been constructed taking into account solid-liquid phase transitions. Data from meteorological stations were used to determine the model parameters, with the help of which the necessary physical and thermophysical characteristics of the computational domain were obtained. For the numerical solution of the problem, the finite volume method (FVM) was used. In order to analyze changes in the temperature field of ice and determine the time required to reach a non-stationary periodic regime, graphs of temperature versus depth were plotted for January at two stations. The study of the results showed that it takes about 50 years of modeling with constant initial data for the temperature of an ice layer up to 20 m deep to reach the periodic regime. For the obtained periodic regime, the temperature versus depth dependences for each month were plotted, and the depth of the active layer, as well as the depth of zero annual amplitudes were found for each meteorological station. A forecast of the ice temperature regime for 2100 was modeled for three Representative Concentration Pathway (RCP) scenarios of global warming: moderate RCP2.6, corresponding to the current emissions of RCP7 and adopted at the Paris Agreement in 2015 RCP1.9. The scenarios are based on the IPCC AR5 and SSP databases, as well as on the existing policy frameworks and declared political intentions of The IEA Stated Policies Scenario (STEPS). The plotted graphs clearly demonstrated that even a moderate RCP2.6 scenario (warming by 2°C) can lead to a serious reduction in the ice cover area, while the RCP7 scenario will lead to even more unsatisfactory consequences. In turn, the scenario of limiting climate warming to 1,5 ° C from pre-industrial levels (RCP1.9) would significantly slow down ice thawing. By analyzing the impact of an additional 0,5°C of warming on other areas, a reduction in the full range of risks to humanity and the planet as a whole becomes evident with the proper efforts of the global community. Thus, the conducted modeling has confirmed the need to reduce the rate of global warming.

## I. Introduction

Modeling the temperature distribution at the poles is a hot topic of the 21st century. Everyone knows that under the influence of global warming, along with permafrost, glaciers and ice begin to thaw, which in turn raises the level of the world's oceans. Decreasing ice surface area also increases the amount of heat absorbed by the ocean. Warming increases risks in many areas of human life, as well as natural risks associated with impacts on biodiversity and ecosystems [1]. As a result, humanity needs to constantly adapt to the ongoing changes.

The poles of the Earth deserve special attention, as they are the points of accumulation of the main mass of ice and cold air masses [2]. Due to the fact that the poles are dominated by ice, it is necessary to consider and model the impact of climate change on the non-stationary periodic ice regime. For this purpose, a model was built for predicting the effect of warming on the ice column at selected meteorological stations.

## II. Problem statement

It is required to numerically simulate the temperature regime in a medium with phase transitions - solid-liquid. Such a state of the medium in a non-stationary one-dimensional formulation is described by the following heat conduction equation:

$$(c\rho + Q\delta(u-u^*))\frac{\partial u}{\partial t} = \frac{\partial}{\partial z}\left(\lambda\frac{\partial u}{\partial z}\right), \qquad (1)$$

where $c$ - specific heat capacity; $\rho$ - density; $\lambda$ - coefficient of thermal conductivity; $u(z,t)$ - temperature of medium; $u^*$ - phase transition temperature; $Q$ - heat of the phase transition; $\delta(u-u^*)$ - delta function.

The solution $u(z,t)$ is to be found in a bounded domain $D=\{0 \le z \le zL\}$, that satisfies the initial condition

$$u(z,0)=\varphi(z). \quad (2)$$

At the upper boundary $z=0$ with temperature $u(0,t)$ convective heat exchange occurs with a medium having a temperature: $\theta(t)$:

$$J = h \cdot (\theta(t) - u(0,t)), \quad (3)$$

where $J$ - heat flow density at the boundary, $h$ – heat transfer coefficient [3].

At the lower boundary $z = zL$ no heat flow condition is set

$$J_b = 0. \quad (4)$$

## III. Physical and geographical conditions

### Geographical location and meteorological data

In problem (1)-(4), the calculation region starts at the ice surface (from the boundary with the atmosphere) and ends in the ice column at a certain depth. The calculations assumed that the heat flow from the Earth's interior would not have a significant effect on the temperature distribution at the selected depth.

To set the upper boundary condition, Table 1 was compiled of mean long-term monthly air temperatures recorded by the following meteorological stations: Amundsen-Scott (WMO Index: 89009) for 1971-2021, Esperanza (WMO Index: 88963) for 1945-2021 in the Antarctic and Cape Morris Jesup (WMO Index 43010) for 1981-2021, Ernst Krenkel Observatory (WMO Index 20046) for 1957-2021 in the Arctic [4-13].

When opened, Amundsen-Scott South Pole Station was located exactly at the South Pole, but at the beginning of 2006, due to ice movements, the station was about 100 meters from the geographic South Pole. The station is located at the coordinates 90°S, 0°E. and at an altitude of 2835 meters above sea level, on a glacier that reaches a maximum thickness of 2850 m nearby.

The Esperanza Base is located at the coordinates 63°24′S, 56°59′W and at an altitude of 24 meters above sea level on the coast of the ocean and is the northernmost point of the continent. This station is also one of two civilian settlements in Antarctica.

The locations of the respective stations in Antarctica are shown in Figure 1.

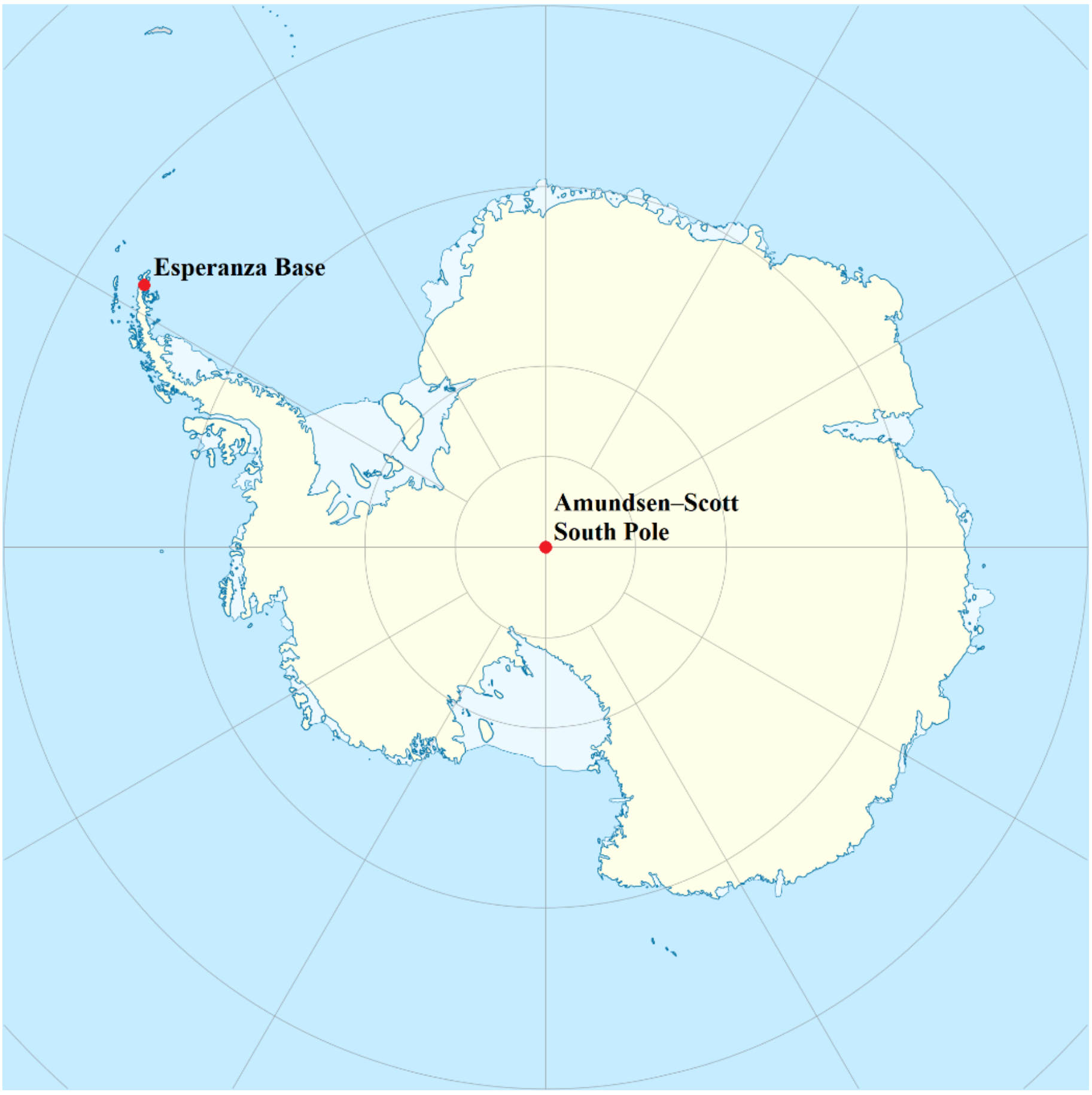

Fig. 1. Location of stations in the Antarctic.

The station at Cape Morris Jesup is located in Peary Land in Greenland at a distance of 709 km from the North Pole at the coordinates 83°37'N, 33°22'W, 4 m above sea level.

The Ernst Krenkel Observatory is the northernmost meteorological station in Russia. The station is located in the north-eastern part of the Heiss island of the Franz Josef Land on the Observatory Cape at the coordinates 80°37'N, 58°3'E. The observatory buildings are located at an altitude of 22 meters above sea level.

The locations of the respective stations in the Arctic are shown in Figure 2.

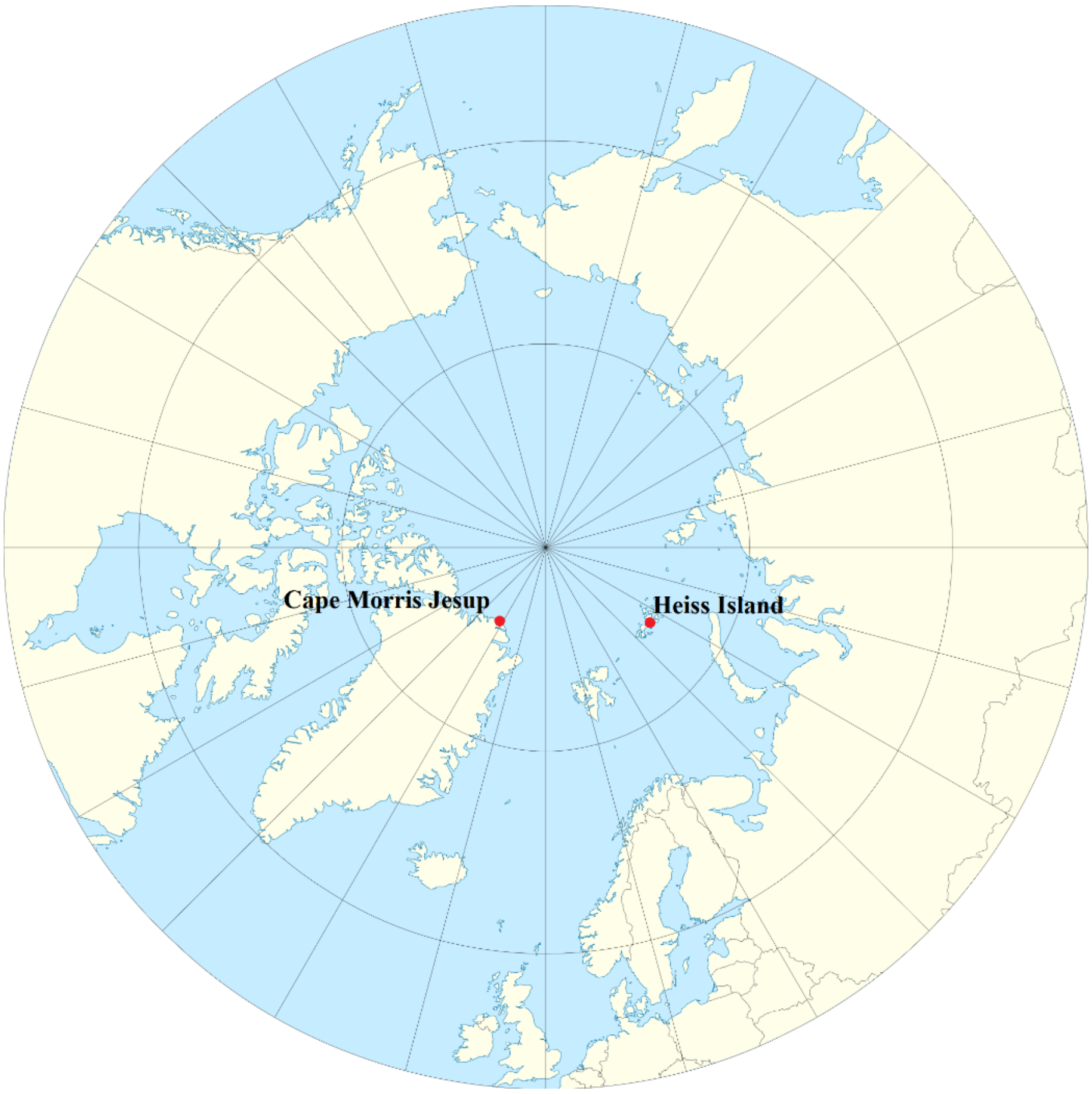

Fig. 2. Location of stations in the Arctic

Table 1 and Table 2 present long-term monthly averages of the parameters required to calculate the thermophysical characteristics of ice and snow cover [4-13].

Table 1. Average long-term values of temperature, wind speed, and snow cover thickness in the Antarctic.

| Amundsen-Scott South Pole Station (WMO 89009) | | | | | | | | | | | | |
|---|---|---|---|---|---|---|---|---|---|---|---|---|
| Month | Jan | Feb | Mar | Apr | May | Jun | Jul | Aug | Sep | Oct | Nov | Dec |
| Temp. | -28 | -40,6 | -53,8 | -57,4 | -57,8 | -58,2 | -60 | -59,7 | -59,1 | -51,1 | -37,9 | -27,5 |
| Snow(cm) | 0 | | | | | | | | | | | |
| Wind speed (m/s) | 4,15 | 4,42 | 5,14 | 5,39 | 5,20 | 5,36 | 5,49 | 5,58 | 5,48 | 5,38 | 5,05 | 4,14 |
| Esperanza Base (WMO 88963) | | | | | | | | | | | | |
| Temp. | 0,8 | -0,1 | -2,6 | -6,5 | -8,8 | -10,8 | -11 | -10 | -7 | -4,1 | -1,6 | 0,4 |
| Snow(cm) | 3,60 | 8,04 | 9,61 | 19,57 | 27,33 | 25,25 | 21,03 | 10,12 | 9,09 | 10,8 | 6,17 | 6,47 |
| Wind speed (m/s) | 5,34 | 6,58 | 7,25 | 7,79 | 7,42 | 8,15 | 8,37 | 8,12 | 8,67 | 8,10 | 6,94 | 5,97 |

Table 2. Average long-term values of temperature, wind speed, and snow cover thickness in the Arctic.

| Cape Morris Jesup (WMO 04301) | | | | | | | | | | | | |
|---|---|---|---|---|---|---|---|---|---|---|---|---|
| Month | Jan | Feb | Mar | Apr | May | Jun | Jul | Aug | Sep | Oct | Nov | Dec |
| Temp. | -30,7 | -31,1 | -31,4 | -23,3 | -10,2 | -0,6 | 1,9 | 0,07 | -8,7 | -18,4 | -25,4 | -28,7 |
| Snow(cm) | 25 | | | | | | | | | | | |
| Wind speed (m/s) | 4,9 | 4,9 | 5,1 | 5,3 | 5 | 4,9 | 4,7 | 4,4 | 4,5 | 4,6 | 4,9 | 5,2 |
| Ernst Krenkel Observatory (WMO 20046) | | | | | | | | | | | | |
| Temp. | -23 | -23,1 | -23,2 | -18,3 | -9 | -1,4 | 0,77 | 0,18 | -2,6 | -10,3 | -16,4 | -21 |
| Snow(cm) | 26,5 | 33,9 | 39,8 | 44,3 | 47,38 | 40,38 | 8,9 | 0,8 | 2,4 | 7,6 | 13,1 | 18,8 |
| Wind speed (m/s) | 6,4 | 5,9 | 5,8 | 5,7 | 5,3 | 5,1 | 4,7 | 4,7 | 6 | 6,5 | 6,2 | 6,1 |

## Physical and thermophysical characteristics of ice

The heat transfer coefficient is calculated by the formula [1]

$$h = \frac{1}{\frac{1}{\alpha} + R}, \tag{5}$$

where $\alpha$ – heat transfer coefficient from ice surface to air;

$R$ – thermal resistance of snow cover.

The heat transfer coefficient $\alpha$ is determined according to an empirical formula for its estimation, which is used in practice [14]

$$\alpha = 5{,}8\sqrt{\omega + 0{,}3}, \tag{6}$$

where $\omega$ – mean monthly wind speed at the surface of the snow cover, taken from meteorological data [3-13].

The thermal resistance of the snow cover is calculated by the formula [15]

$$R = \frac{d_s}{\lambda_s}, \tag{7}$$

where $d_s$ – average monthly snow depth, based on meteorological data [3-13];

$\lambda_s$ – the average monthly thermal conductivity of the snow cover, that is determined by the formula [15]

$$\lambda_s = m_d(0{,}18 + 0{,}87\rho_s), \tag{8}$$

where $m_d = 1\ kcal/(t \cdot m \cdot h \cdot {}^\circ C)$ – conversion factor for our case;

$\rho_s$ – average monthly snow cover density, $t/m^3$, based on meteorological data [7-13].

In areas with an average wind speed in the winter period over 5 $m/s$, calculated by the formula (6) the value of $R$ should be increased by 1,3 times [15].

Using the data for $d_s$, $\rho_s$, $\omega$ and formulas (5)-(8) Table 3 and Table 4 were compiled, which reflect the physical characteristics at the upper boundary of the computational domain.

Table 3. Characteristics of ice at the upper boundary of the calculation domain in the Antarctic.

| Amundsen-Scott South Pole Station | | | | | | | | | | | | |
|---|---|---|---|---|---|---|---|---|---|---|---|---|
| Month | 1 | 2 | 3 | 4 | 5 | 6 | 7 | 8 | 9 | 10 | 11 | 12 |
| $\alpha, \frac{kcal}{m^2 h ^\circ C}$ | 12,24 | 12,6 | 13,53 | 13,83 | 13,6 | 13,8 | 13,96 | 14,06 | 13,94 | 13,83 | 13,41 | 12,22 |
| $\rho_s, kg / m^3$ | N/A | | | | | | | | | | | |
| $R, \frac{m^2 h ^\circ C}{kcal}$ | N/A | | | | | | | | | | | |
| $h, \frac{kcal}{m^2 h ^\circ C}$ | 12,24 | 12,61 | 13,53 | 13,83 | 13,6 | 13,8 | 13,96 | 14,1 | 13,94 | 13,83 | 13,4 | 12,22 |
| Esperanza Base | | | | | | | | | | | | |
| $\alpha, \frac{kcal}{m^2 h ^\circ C}$ | 13,77 | 15,22 | 15,94 | 16,5 | 16,11 | 16,86 | 17,08 | 16,83 | 17,38 | 16,81 | 15,61 | 14,53 |
| $\rho_s, кг / м^3$ | 500 | | | | | | | | | | | |
| $\lambda_s, \frac{kcal}{mh ^\circ C}$ | 0,615 | | | | | | | | | | | |
| $R, \frac{m^2 h ^\circ C}{kcal}$ | 0,08 | 0,17 | 0,2 | 0,41 | 0,58 | 0,53 | 0,44 | 0,21 | 0,19 | 0,23 | 0,13 | 0,14 |
| $h, \frac{kcal}{m^2 h ^\circ C}$ | 6,72 | 4,24 | 3,76 | 2,11 | 1,56 | 1,69 | 1,99 | 3,66 | 4 | 3,47 | 5,14 | 4,86 |

Table 4. Characteristics of ice at the upper boundary of the calculation domain in the Arctic.

| Cape Morris Jesup | | | | | | | | | | | | |
|---|---|---|---|---|---|---|---|---|---|---|---|---|
| Month | 1 | 2 | 3 | 4 | 5 | 6 | 7 | 8 | 9 | 10 | 11 | 12 |
| $\alpha, \frac{kcal}{m^2 h ^\circ C}$ | 13,23 | 13,25 | 13,5 | 13,7 | 13,4 | 13,24 | 13 | 12,6 | 12,7 | 12,8 | 13,2 | 13,6 |
| $\rho_s, kg / m^3$ | 264,3 | | | | | | | | | | | |
| $R, \frac{m^2 h ^\circ C}{kcal}$ | 7,93 | | | | | | | | | | | |
| $\lambda_s, \frac{kcal}{mh ^\circ C}$ | 0,41 | | | | | | | | | | | |
| $h, \frac{kcal}{m^2 h ^\circ C}$ | 1,455 | | | | | | | | | | | |
| Ernst Krenkel Observatory | | | | | | | | | | | | |
| $\alpha, \frac{kcal}{m^2 h ^\circ C}$ | 15,01 | 14,44 | 14,32 | 14,21 | 13,73 | 13,48 | 12,97 | 12,97 | 14,56 | 15,12 | 14,79 | 14,67 |
| $\rho_s, kg / m^3$ | 400 | | | | | | | | | | | |
| $\lambda_s, \frac{kcal}{mh ^\circ C}$ | 0,528 | | | | | | | | | | | |
| $R, \frac{m^2 h ^\circ C}{kcal}$ | 0,65 | 0,84 | 0,98 | 1,09 | 1,17 | 0,99 | 0,22 | 0,02 | 0,06 | 0,19 | 0,32 | 0,46 |
| $h, \frac{kcal}{m^2 h ^\circ C}$ | 1,39 | 1,1 | 0,95 | 0,86 | 0,81 | 0,94 | 3,35 | 10,3 | 7,79 | 3,96 | 2,56 | 1,88 |

The density of pure freshwater ice, devoid of any pores, gas inclusions and impurities at temperature 0°C and atmospheric pressure 1000 mbar is equal to 916.8 kg / m3 [16]. As the temperature decreases, the density due to compression increases and can be calculated by the formula

$$\rho_{iT} = \frac{\rho_0}{1 + \gamma \cdot T}, kg / m^3, \quad (9)$$

where $\rho_0$ – density of ice without cavities at temperature 0 °C;

$T$ – ice temperature;

$\gamma$ – volumetric thermal expansion coefficient of pure ice, average value $\gamma = 1{,}58 \cdot 10^{-4} K^{-1}$.

For water and ice at 0 °C and atmospheric pressure of 101.3 kPa, the values of molar heat capacities are equal to 75,3 and 37,7 J/(mol-K) respectively, i.e., the heat capacity of ice is about half of water.

To calculate the specific heat capacity of freshwater ice at normal atmospheric pressure with decreasing temperature, the following law was used, derived from the empirical data of Dickinson and Osborn [17].

$$c_i = (2{,}114 + 0{,}007787 \cdot T) \cdot 0{,}2388,\, kcal / \left(kg \cdot^{\circ} C\right), \quad (10)$$

where $T$ – ice temperature.

The average thermal conductivity of freshwater ice near the melting point at normal atmospheric pressure is four times greater than that of freshwater at 0°C.

Temperature dependences of the thermal conductivity coefficient of polycrystalline freshwater ice at normal atmospheric pressure and at temperatures from 0 to -130°C follow the law proposed by Yu.A. Nazintsev on the basis of published data and his own experiments [17]:

$$\lambda_{iT} = 2{,}24 \cdot (1 - 0{,}0048 \cdot T),\, kcal / \left(m \cdot h \cdot^{\circ} C\right), \quad (11)$$

where $T$ – ice temperature.

The dependence of freshwater ice thermal conductivity coefficient on density within the usual density of natural freshwater ice (800 - 910 kg / m3) can be assumed to be linear [17]

$$\lambda_{\rho T} = \lambda_{iT} - 0{,}0057(\rho_0 - \rho_{iT}) \cdot 0{,}86,\, kcal / \left(m \cdot h \cdot^{\circ} C\right), \quad (12)$$

where $\lambda_{iT}$ – thermal conductivity coefficient of ice without cavities at temperature $T$;

$\rho_0$ – density of ice without cavities at 0°C;

$\rho_{iT}$ – density of ice without cavities at temperature $T$.

Water characteristics also tend to change with temperature.

Water density $\rho_w \approx 1000,\, kg / m^3$ in the temperature range 0-10°C. With a further increase in temperature, the density can be calculated using the following approximate formula [18], [19]

$$\rho_w = \frac{995,7}{0,984 + 0,483 \cdot 10^{-3} \cdot T}, kg / m^3, \quad (13)$$

where $T$ – water temperature.

The heat capacity of water $c_w = 1,007\ kcal / (kg \cdot ^\circ C)$ for temperatures below 10°C. With increasing temperature, an approximate formula is used [18], [19]

$$c_w = (4194 - 1,15 \cdot T + 1,5 \cdot 10^{-2} \cdot T^2) \cdot 0,2388, kcal / (kg \cdot ^\circ C), \quad (14)$$

where $T$ – water temperature.

Thermal conductivity of water $\lambda_w = 0,489\ kcal / \left(m \cdot h \cdot ^\circ C\right)$ for temperatures below 10°C. With increasing temperature, an approximate formula is used [18], [19]

$$\lambda_w = 0,553 \cdot (1 + 0,003 \cdot T) \cdot 0,86, kcal / \left(m \cdot h \cdot ^\circ C\right), \quad (15)$$

where $T$ – water temperature.

The value of volumetric heat of freezing of water (ice melting) is taken as equal to the amount of heat necessary for freezing of water (ice melting) in a unit volume of ground and determined by the formula [15]

$$Q = \kappa \rho_i, \quad (16)$$

where $\kappa = 79,4\ \ kcal / kg$ – specific heat of water-ice phase transition.

The formula for the volumetric heat capacity, which takes into account phase transitions, has the form

$$c\rho = \begin{cases} \rho_{iT} c_{iT}, & u < u^*; \\ \rho_w c_w, & u > u^*, \end{cases} \quad (17)$$

where temperature of phase transition $u^* = 0$ °C.

## IV. Numerical solution

For numerical solution of the problem with phase transitions, the computational algorithm is constructed without explicitly identifying the phase transition boundary. After the solution is found, the phase transition boundary can be found as a surface having the temperature $u = u^*$.

In the approximate solution of the problem (1)-(12), the coefficient in the left part of equation (1) is smoothed and the transition to the usual problem of heat conduction is made [20]. The delta function $\delta(u - u^*)$ is replaced by the function $\delta((u - u^*), \Delta)$, which is nonzero only inside the smoothing interval $[-\Delta, \Delta]$. As a result, instead of solving (1), the solution of the equation with the smoothed coefficient is searched for

$$(c\rho + \mathrm{Q}\delta((u - u^*), \Delta))\frac{\partial u}{\partial t} = \frac{\partial}{\partial z}\left(\lambda \frac{\partial u}{\partial z}\right). \tag{18}$$

For the approximation of delta function the formulas, which are constructed by taking into account the condition of conservation of heat balance on the interval $[-\Delta, \Delta]$, are used. In this paper a step approximation is used

$$\delta(u - u^*, \Delta) = \begin{cases} \dfrac{1}{2\Delta}, & \left|u - u^*\right| \le \Delta, \\ 0, & \left|u - u^*\right| \ge \Delta \end{cases}.$$

It can be seen that for this formula the condition of conservation of heat balance is satisfied:

$$\int_{-\Delta}^{\Delta} \delta((u - u^*), \Delta) du = 1.$$

The value of the smoothing parameter $\Delta$ in the calculations is taken to be two.

Numerical solution of equation (18) with appropriate conditions (2)-(17) is obtained by finite volume method (FVM) [21], [22]. In the final version, numerical calculations were performed with the number of finite volumes equal to 200 and time step $\Delta t = 0{,}001$.

In order to study the temperature regime's transition to a non-stationary periodic process, two stations were selected: Amundsen-Scott, with an initial ice temperature of -40°C, and the Ernst Krenkel Observatory, with an initial ice temperature of -20°C. Simulation for January at each station helped to determine the time after what the temperature regime of 20 m column of ice will fully stabilize.

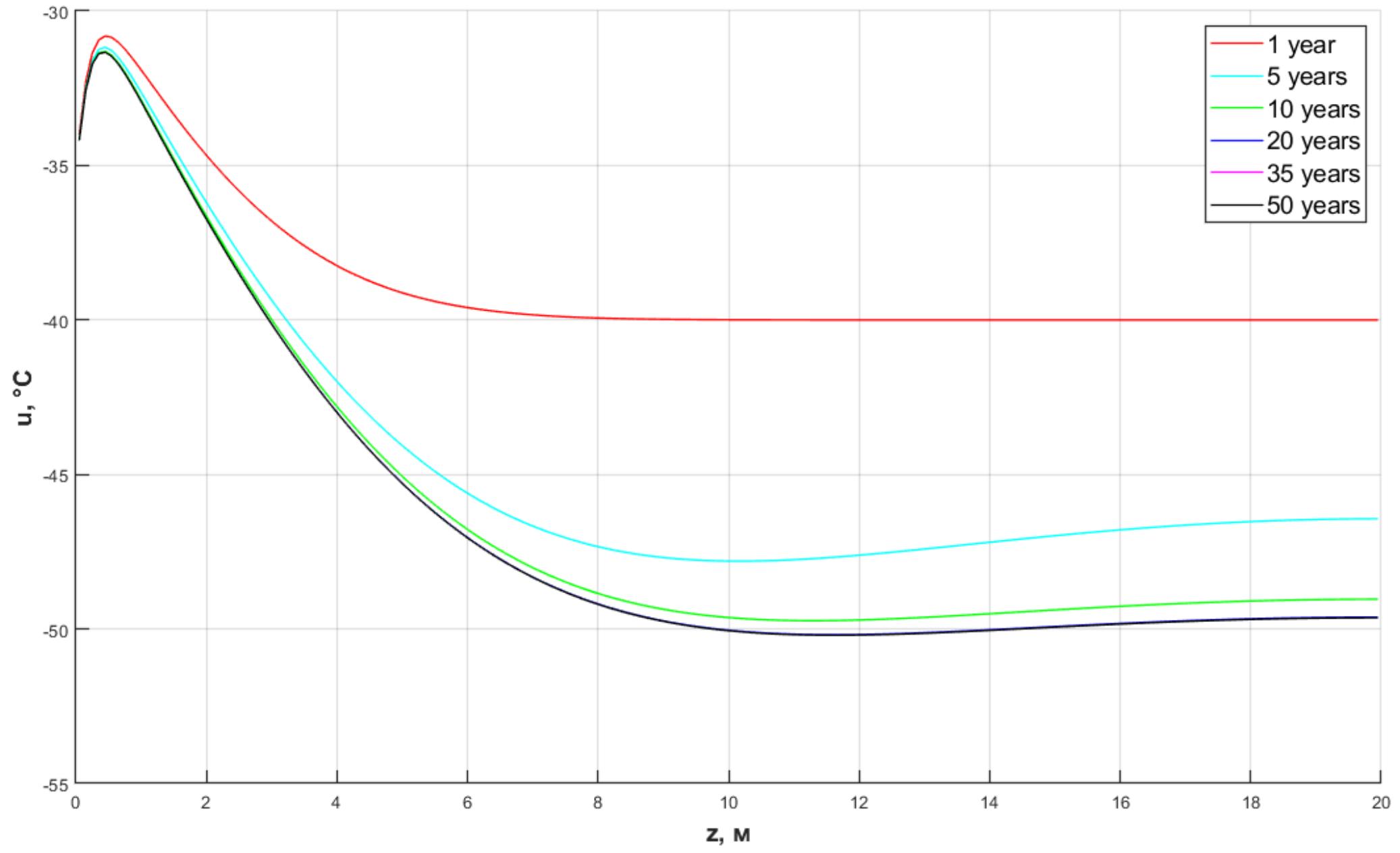

a

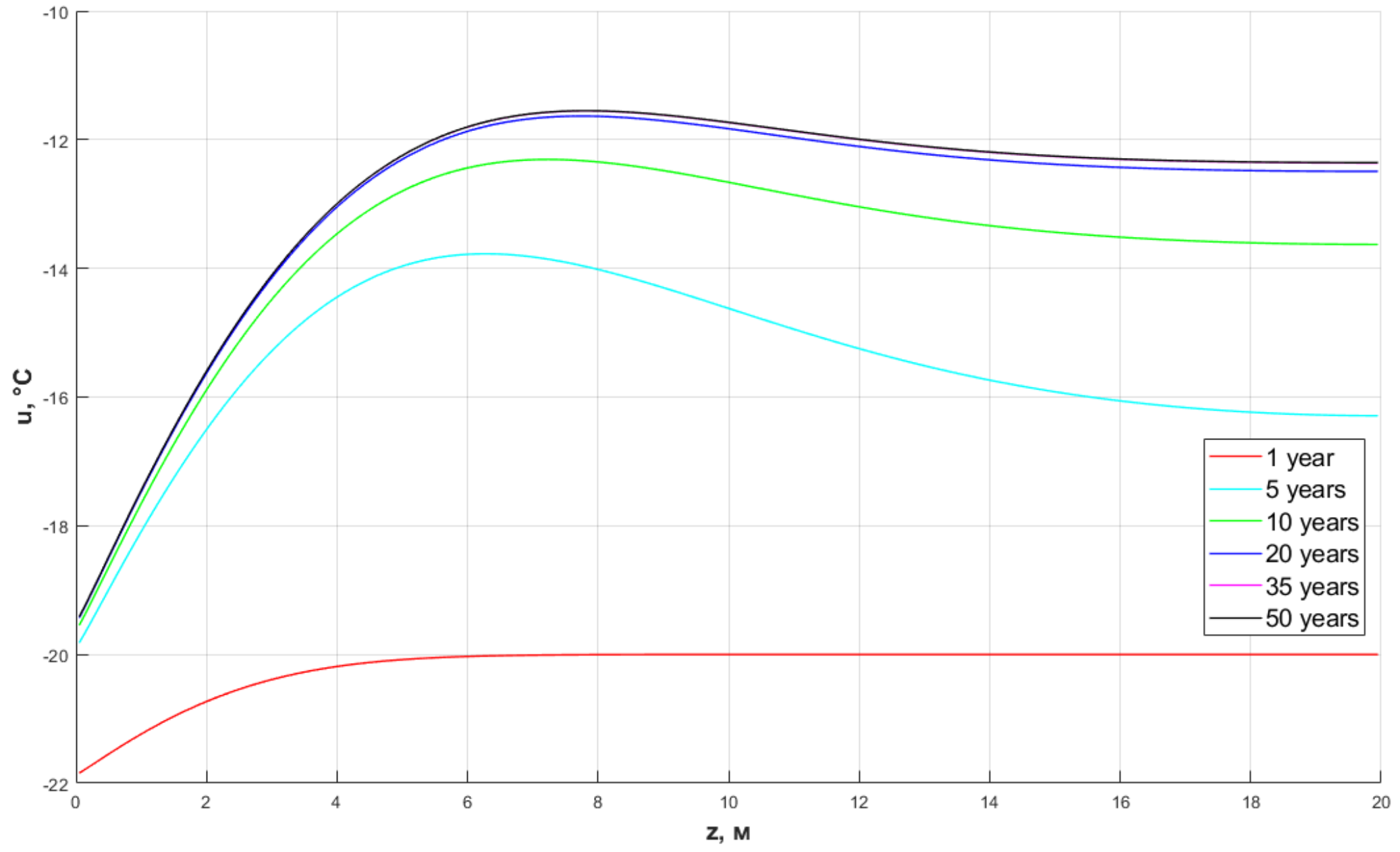

b

Fig. 3. Graphs of ice temperature distribution in January at two stations: (a) Amundsen-Scott (b) E. Krenkel Observatory.

The obtained graphs (Fig. 3) show that after 1 year the temperature distribution in both cases begins to shift from the randomly chosen initial conditions. After 50 years, the graphs begin to coincide with sufficient accuracy, i.e. the dependence of ice temperature on time has reached the non-stationary periodic regime.

To further study the temperature regime of the ice column during the year, non-stationary periodic regimes were plotted for each month at all the selected stations, taking into account the estimated time required for stabilization.

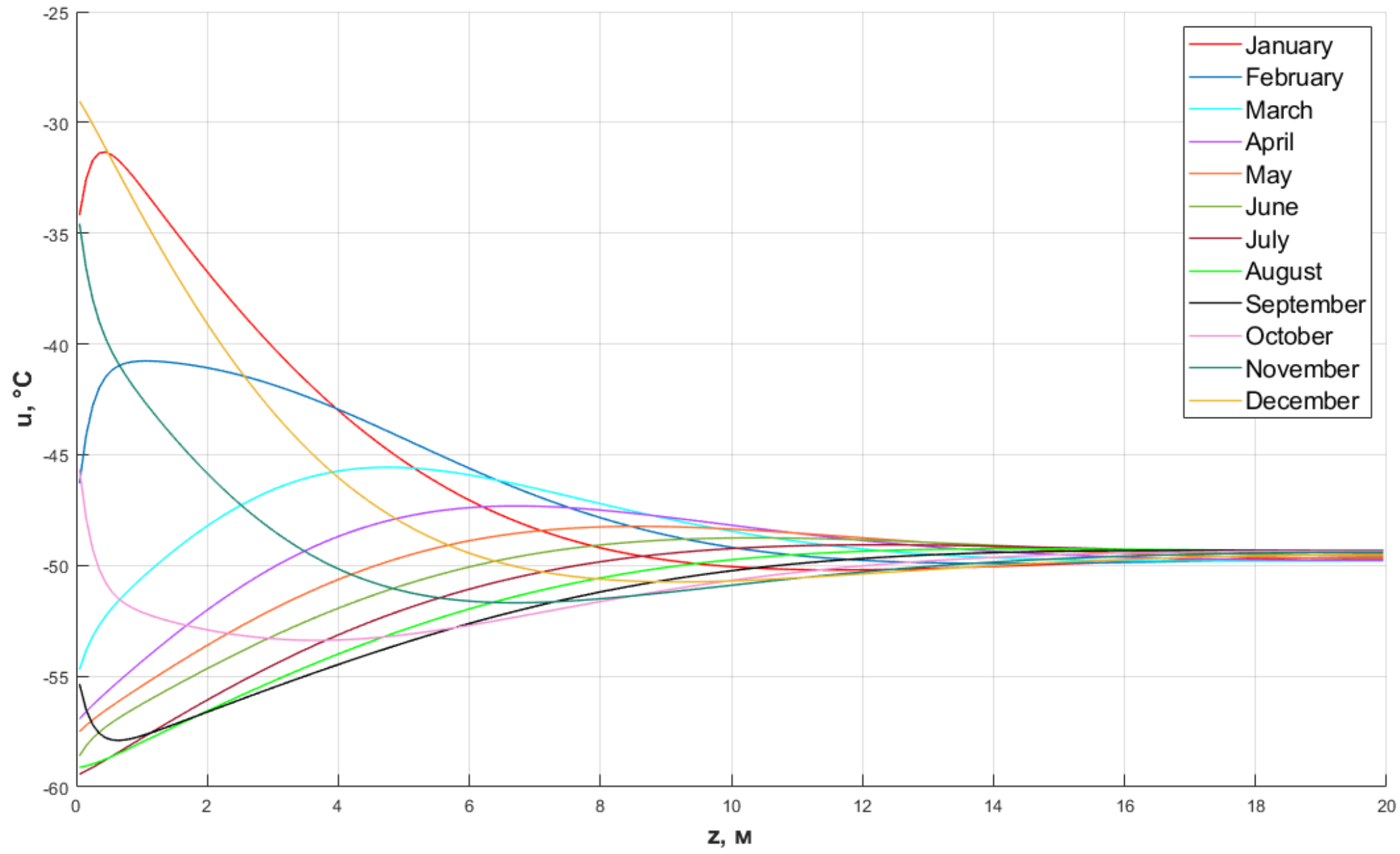

a

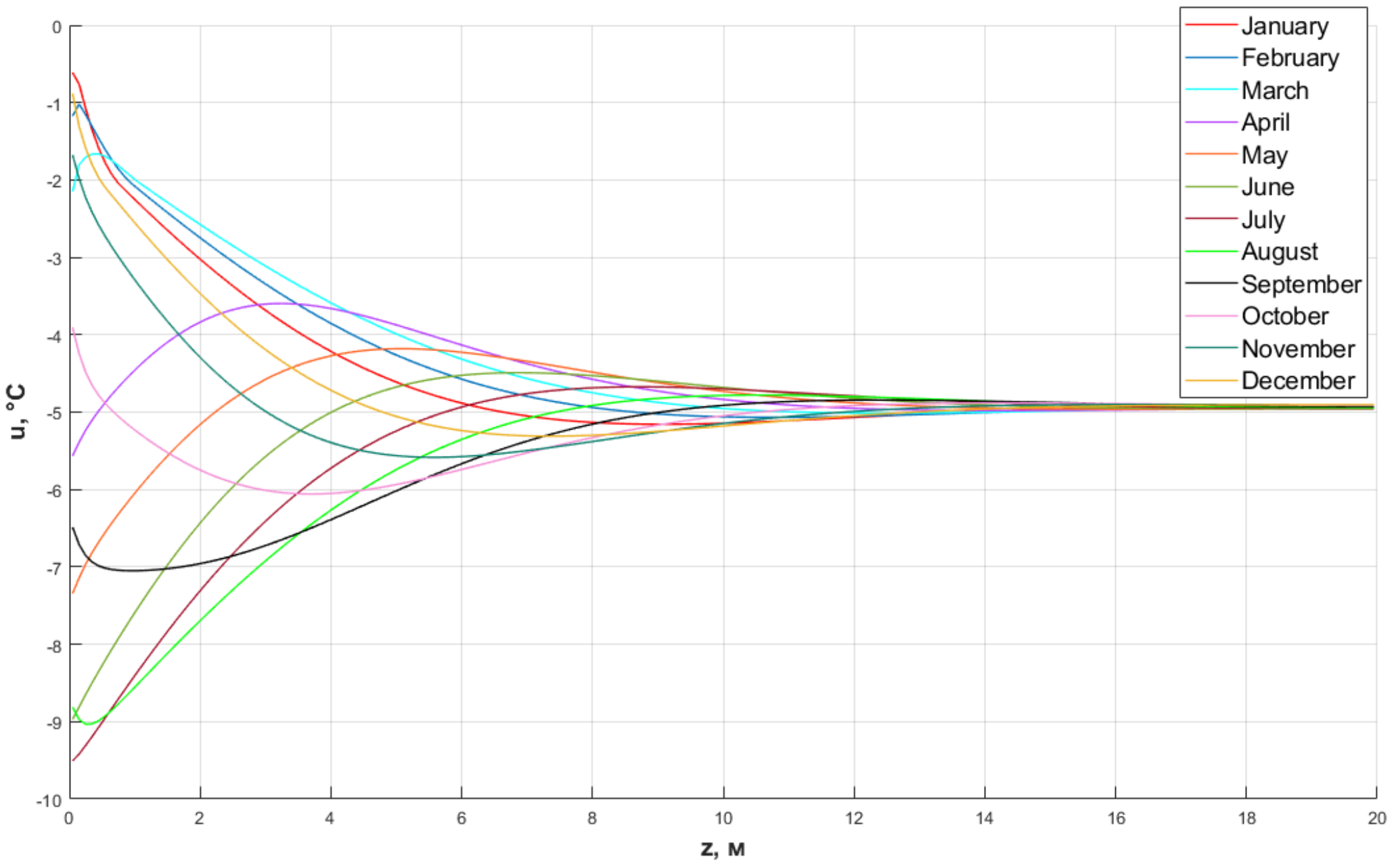

b

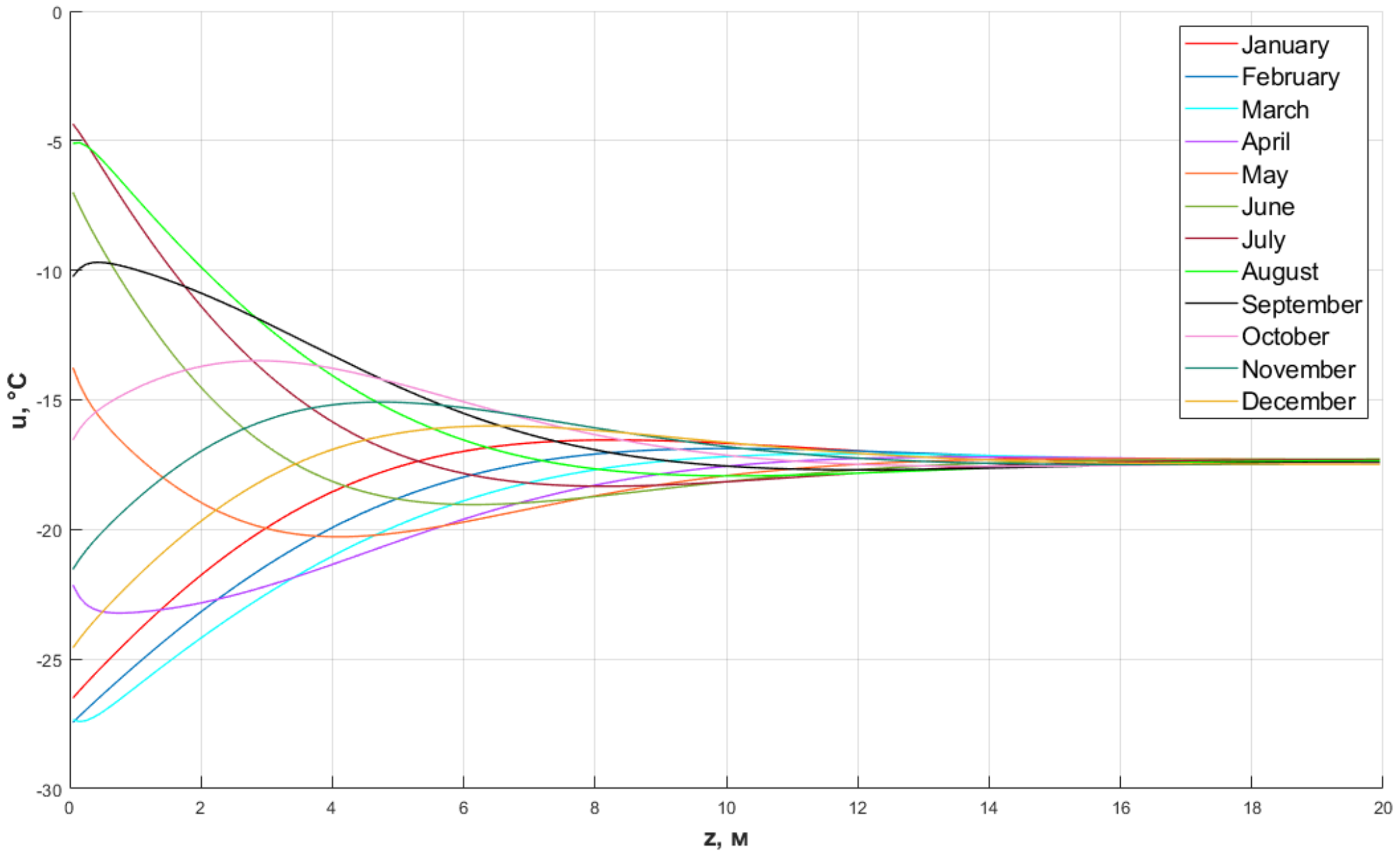

c

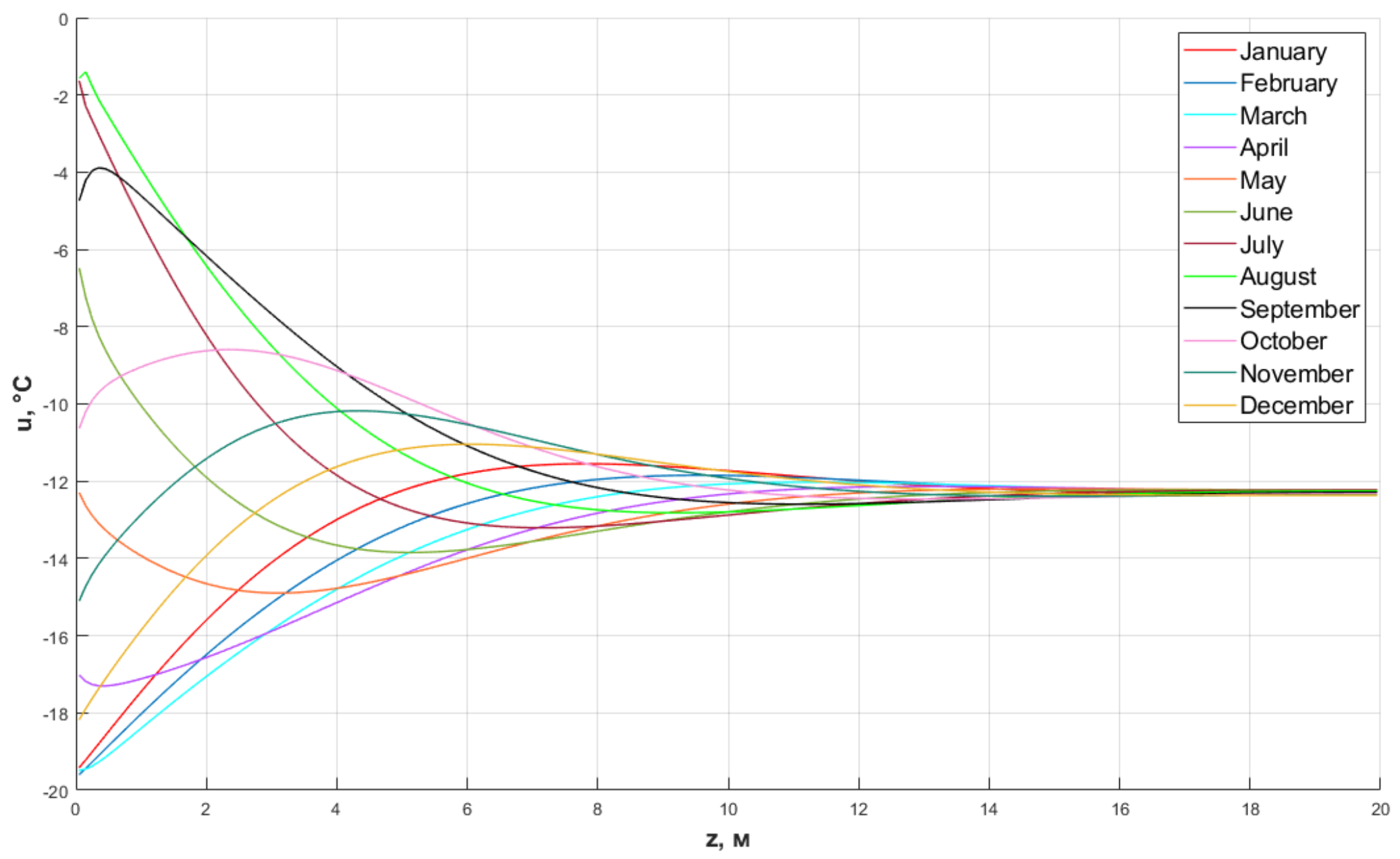

d

Fig. 4. Non-stationary periodic regimes of ice temperature for each month at each station:
(a) Amundsen-Scott, (b) Esperanza, (c) Cape Morris Jesup, (d) Ernst Krenkel Observatory.

The first station in the Antarctic is Amundsen-Scott (Fig. 4, a). The ice temperature near the surface throughout the year varies from -59,4 to -29°C, at a depth of 4 m from -54,5 to -42,9°C. The depth of zero annual amplitudes corresponds to 16 m [23]. With further immersion into the strata, the seasonal temperature fluctuations stabilize and the non-stationary temperature regime tends to -49,5°C, which corresponds to empirical studies with a slight shift in depth, since in real conditions the ground temperature distribution does not have time to reach a similar non-stationary periodic regime due to constant changes [24].

The second station is Esperanza (Fig. 4, b). The ice temperature near the surface varies from -9,5 to -0,6°C, at a depth of 4 m in the range from -6,4 to -3,6°C. The zero annual amplitude depth is 14 m, below this mark the non-stationary ice temperature regime is -4,9°C.

The first station in the Arctic is Cape Morris-Jesup (Fig. 4, c). The ice temperature near the surface varies from -27,4 to -4,4°C, at a depth of 4 m from - 21,4 to -13,3°C. The depth of the zero annual amplitudes corresponds to approximately 14 m, then the non-stationary temperature regime goes to the value of -17,4°C.

The second station is the Ernst Krenkel Observatory (Fig. 4, d). The ice temperature near the surface ranges from -1,6 to -19,6°C, at a depth of 4 m it ranges from -15,2 to -9°C. The depth of zero amplitudes is about 14 m, with further immersion in ice the non-stationary temperature regime tends to -12,3°C.

## V. Prediction of ice column temperature regime with regard to climate warming

The ongoing climate changes, in particular global warming, affect not only the ground, but also the temperature regime of the ice [25]. Two scenarios were chosen to simulate the changes.

The first scenario, RCP2.6, is the "very severe" pathway [26], [28]. According to the IPCC, RCP2.6 requires that carbon dioxide (CO2) emissions begin to decline by 2020 and fall to zero by 2100. It also requires that methane (CH4) emissions fall to about half the CH4 2020 level and that sulfur dioxide (SO2) emissions fall to about 10% of 1980-1990 emissions. Like all other RCPs, RCP2.6 requires negative CO2 emissions (such as absorption by trees). For RCP2.6, these negative emissions would be 2 gigatons of CO2 per year [27]. RCP2.6 is likely to keep the global temperature rise below 2°C by 2100 [27].

The second RCP7 model implies a scenario with preservation of current emissions up to the year 2100 without any mitigation or limitations [26], [28]. In it the increase in the global average temperature will be about 4°C. It was chosen to replace RCP8.5 from past studies because this scenario is considered increasingly unlikely each year and continues to be used either to track historical total cumulative CO2 emissions or shorter-term projections [25], [29] [30], [31].

Warming is above the annual global average in many regions of the globe, over land it is mostly higher compared to the ocean, and there is also variation by time of year. For the numerical estimate of climate change in the Antarctic and the Arctic, the two aforementioned prognostic models of carbon dioxide content (Representative Concentration Pathway, RCP) with station-specific warming values were used. The 1184 AR5(IPCC 5th Assessment Report) [32], [33] and 127 SSP(Shared Socioeconomic Pathways) [34], [35] scenarios were combined for the numerical warming assessment, taking into account projected CO2 emissions growth rates and all current political trends [36], [37], [38].

Table 5. Numerical values of warming scenarios.

| Station | Amundsen-Scott | Esperanza | Morris Jesup | E. Krenkel Observatory |
|---|---|---|---|---|
| RCP2.6, °C | 1,875 | 1,125 | 4,5 | 6 |
| RCP7, °C | 4,5 | 4 | 10 | 12 |

Table 5 shows the numerical values of warming scenarios for each station by 2080-2100 years.

These results show that warming in the Arctic is 2-3 times greater than in the Antarctic [39]. The Ernst Krenkel Observatory is located in the most heated region of the planet, where the warming reaches 12°C in the RCP7 scenario.

In the first step the graphs for the predictive model of changes in the temperature distribution of the ice column for the RCP2.6 scenario were plotted (Fig. 5).

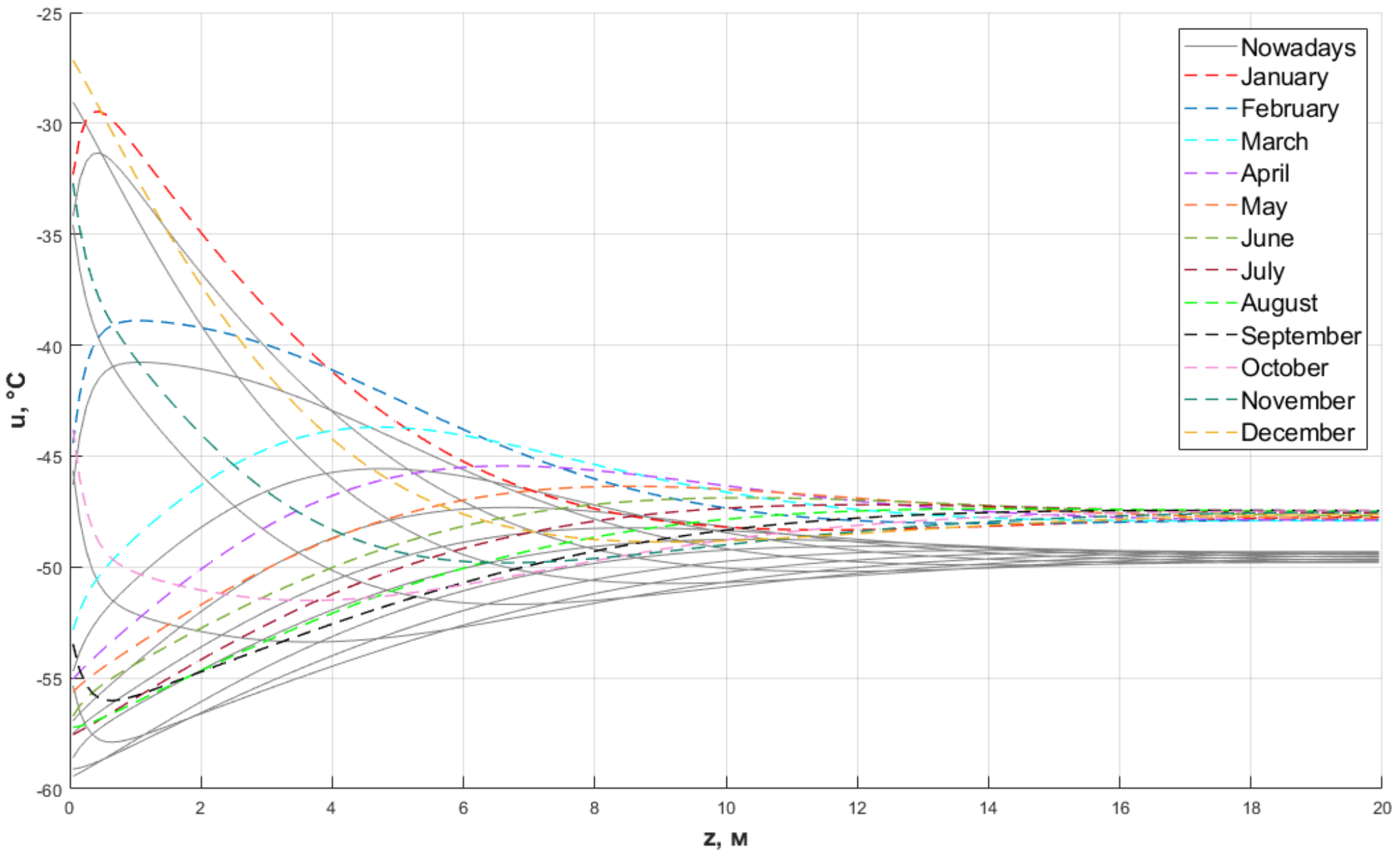

a

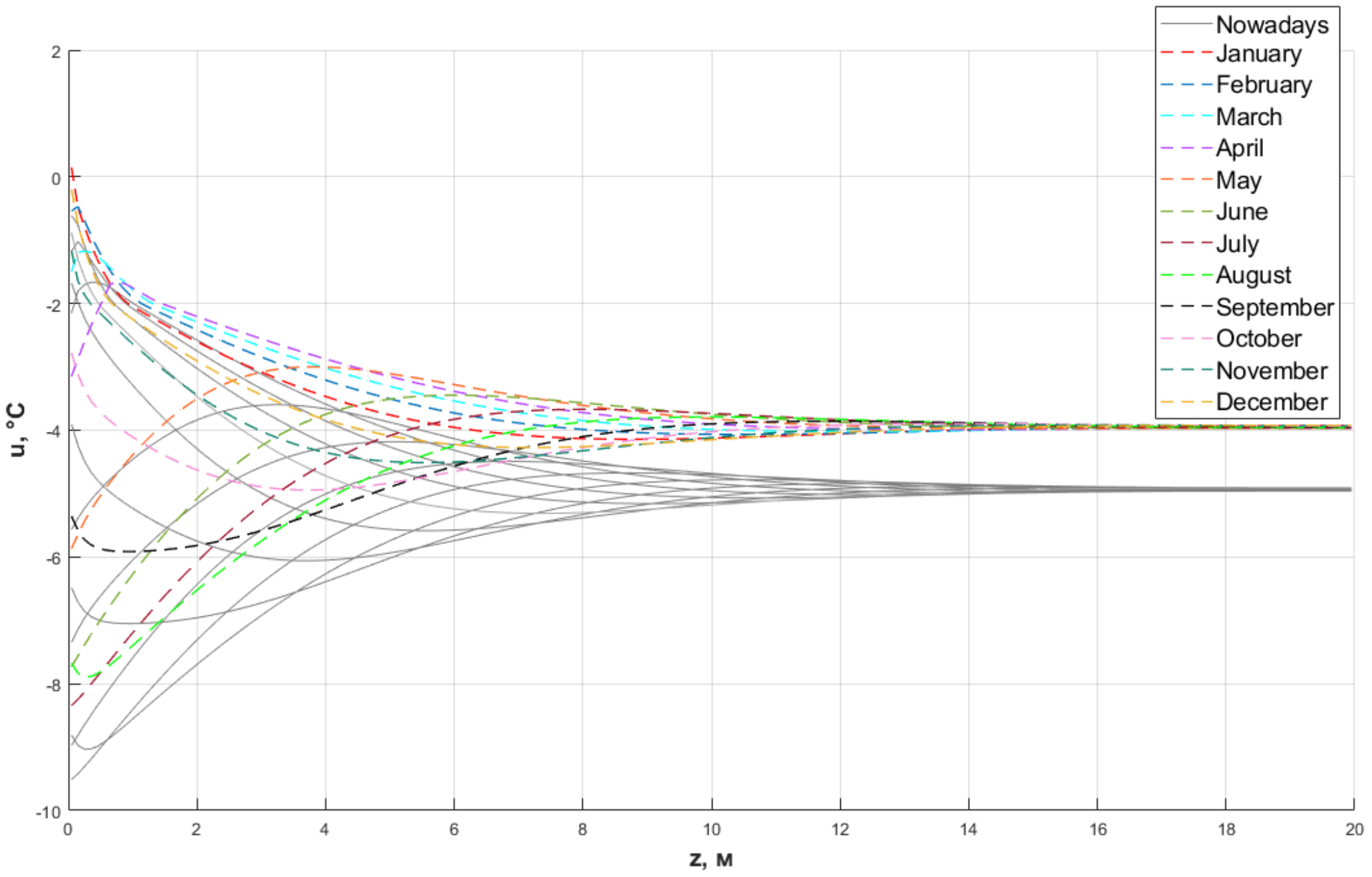

b

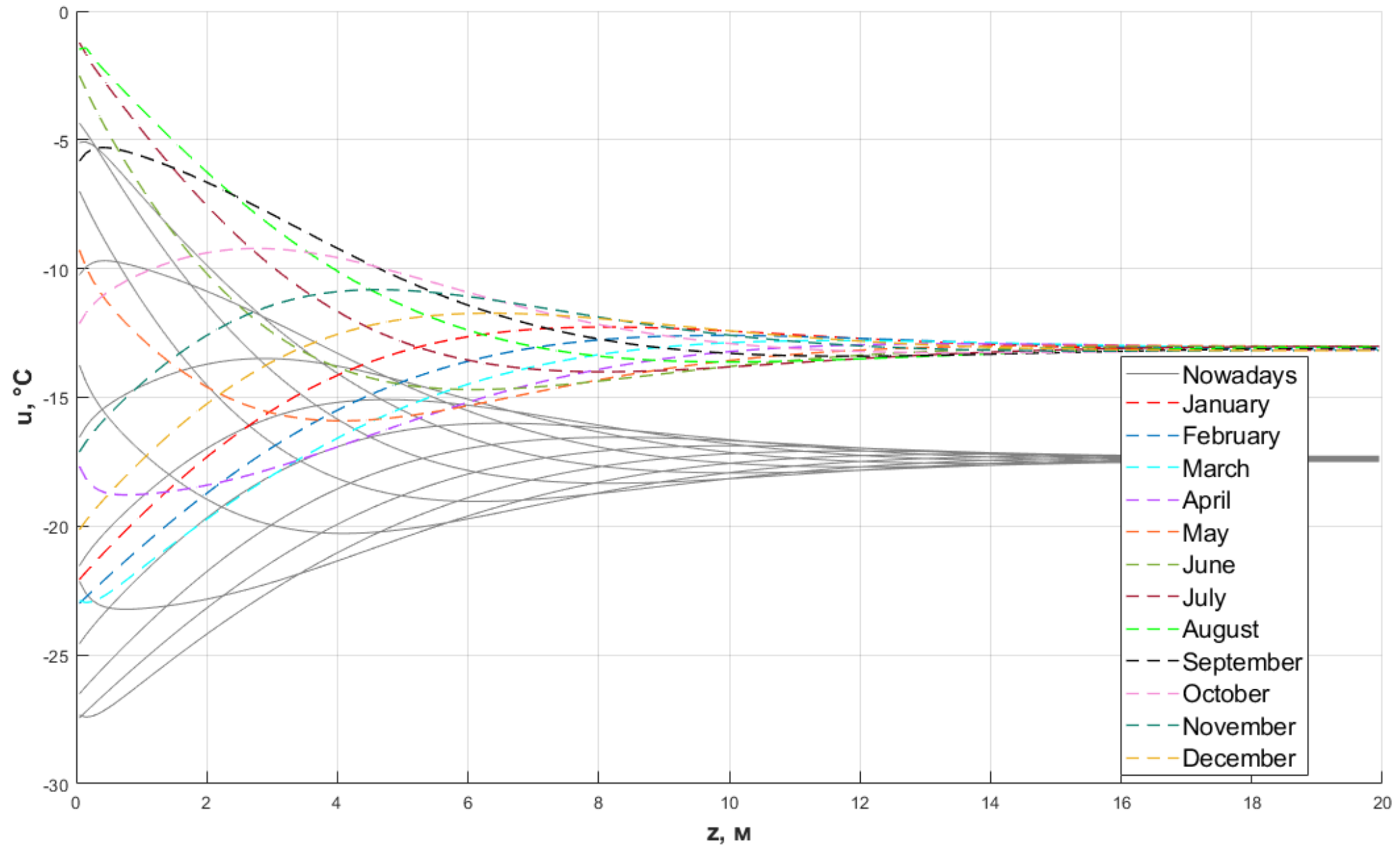

c

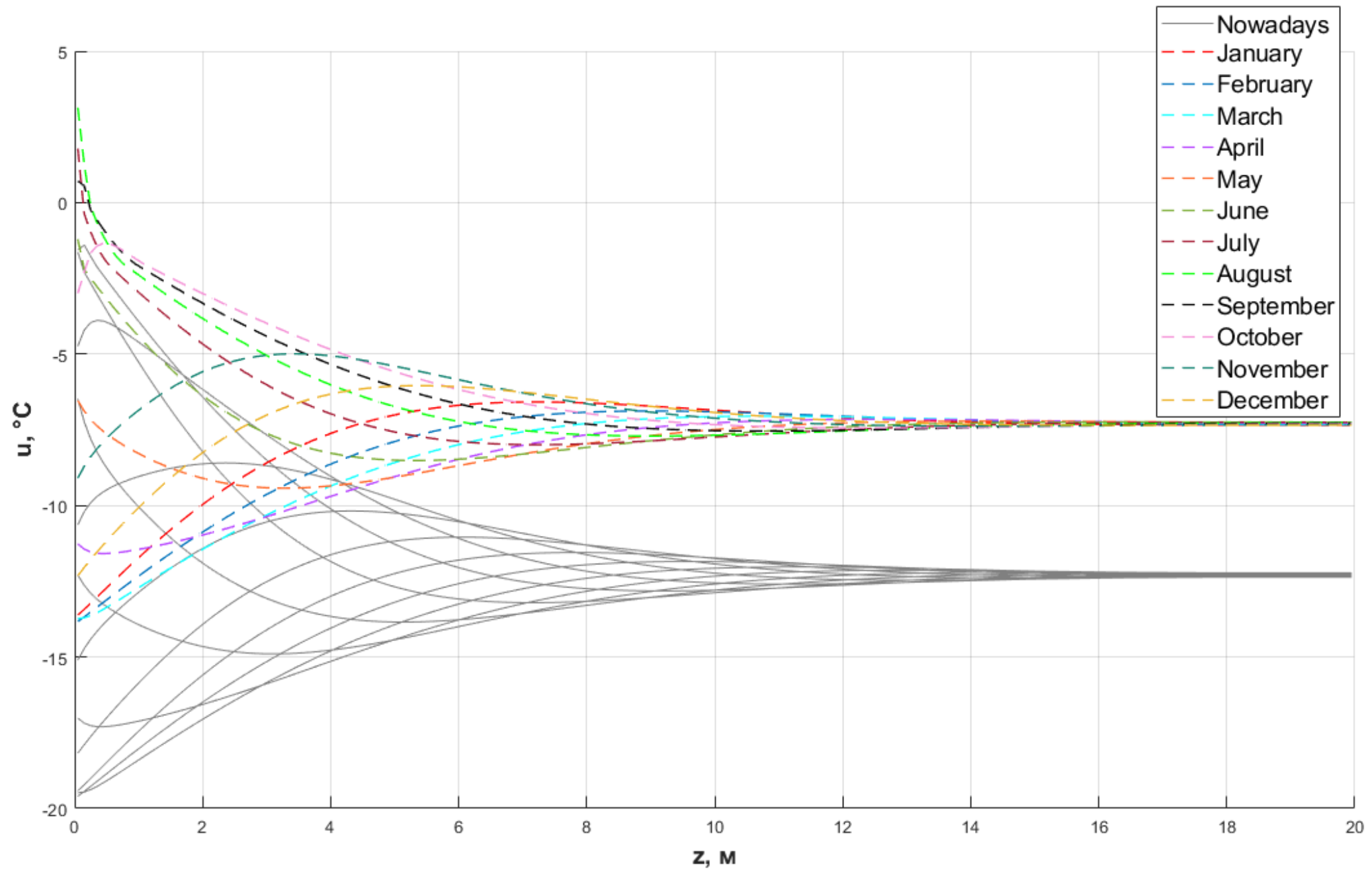

d

Fig. 5. Graphs of non-stationary periodic regimes of ice temperatures nowadays (gray) and predicted for the RCP2.6 scenario for 2080-2100. (colored): (a) Amundsen-Scott, (b) Esperanza, (c) Cape Morris Jesup, (d) Ernst Krenkel Observatory.

In Antarctica, at Amundsen-Scott South Pole Station (Fig. 5, a) near the surface the ice temperature changes in the range from -57,6 to -27,2°C (was -59,4 to -29°C), at a depth of 4 m in the range from -52,6 to -41,1°C (was -54,5 to -42,9°C), the warming atmosphere is reflected on the ice almost linearly due to the absence of snow cover. The depth of the zero annual amplitudes actually remained equal to 16 m. The non-stationary periodic temperature regime shifted by 1.8°C towards warming and reached -47,7°C.

At Esperanza Base (Fig. 5, b) near the surface the ice temperature varies from -8,3 to 0,15°C (was -9,5 to -0,6°C), at 4 m depth it varies from -5,3 to -2,9°C (was -6,4 to -3,6°C). In this scenario, an active layer appears with a depth of 0,08 m, i.e. ice thawing is observed. The depth of the zero annual amplitudes remained approximately equal to 14 m. The non-stationary periodic temperature regime shifted by 1°C towards warmth and reached a value of -3,9°C.

In the Arctic, at Cape Morris Jesup (Fig. 5, c) near the surface the ice temperature changes in the range from -23 to -1,2°C (was -27,4 to -4,4°C), at a depth of 4 m in the range from -16,9 to -9,2°C (was -21,4 to -13,3°C). The depth of zero annual amplitudes remained approximately equal to 14 m. The non-stationary periodic temperature regime has shifted by 4,3°C and reached the value of -13,1°C.

At the Ernst Krenkel Observatory (Fig. 5, d) near the surface the ice temperature varies from -13,8 to 3°C (was -19,6 to -1,6°C), at a depth of 4 m it varies from -9,7 to -4,9°C (was -15,2 to -9°C). This scenario produced an active layer with a depth of 0,24 m. The depth of zero annual amplitudes decreased to 13 m (was 14 m). The non-stationary periodic temperature regime shifted by 5°C and reached -7,3°C.

Simulations performed for the years 2080-2100 of the RCP2.6 scenario demonstrated the appearance of an active layer at the Ernst Krenkel Observatory and Esperanza Base, and at Cape Morris Jesup the ice temperature approached the melting point. Such results point to the need for mankind to strive for a more restrained warming than in this scenario.

In continuation of the study, graphs were plotted for the predictive model of changes in the temperature distribution of the ice column for the RCP7 scenario (Fig. 6)

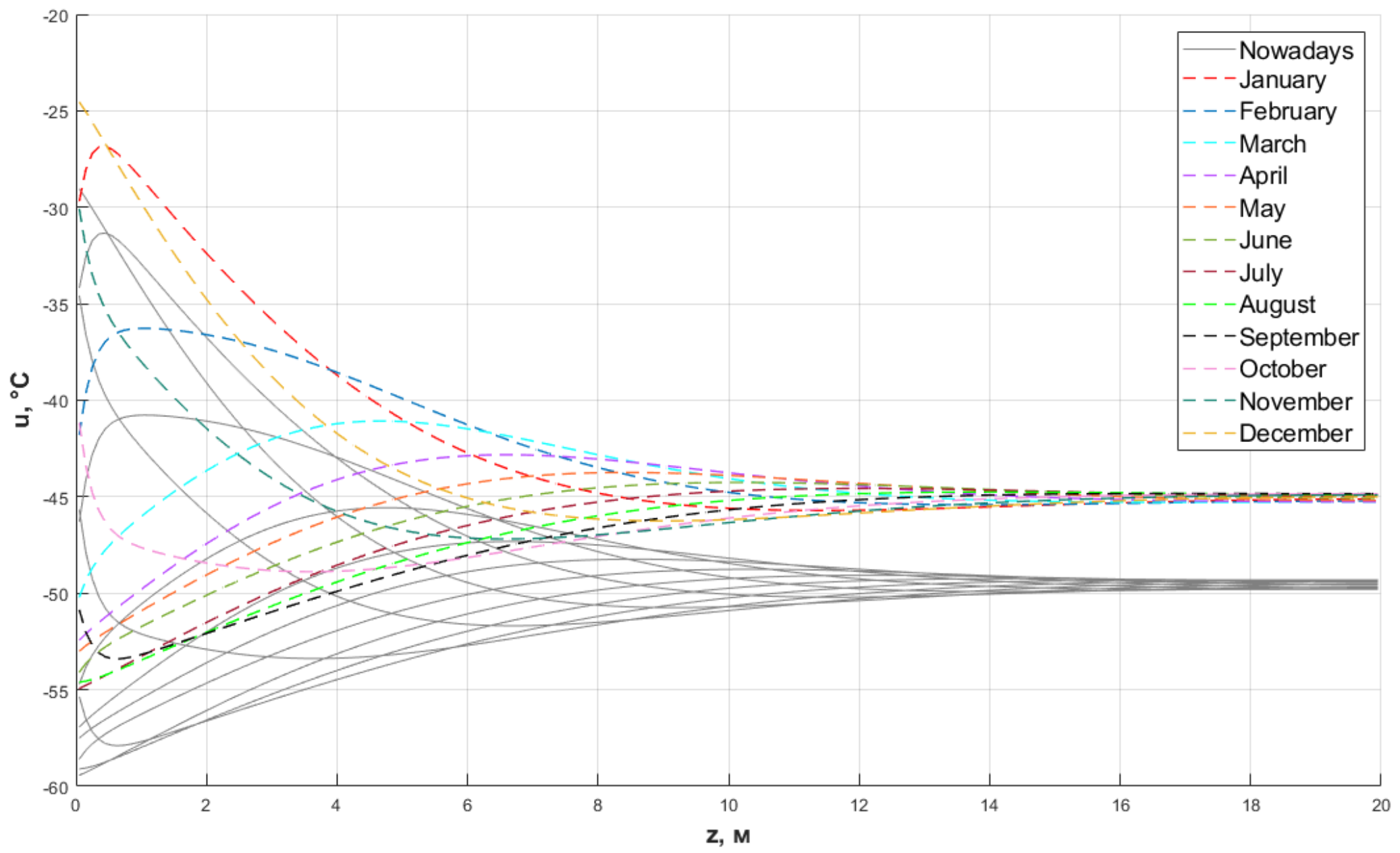

a

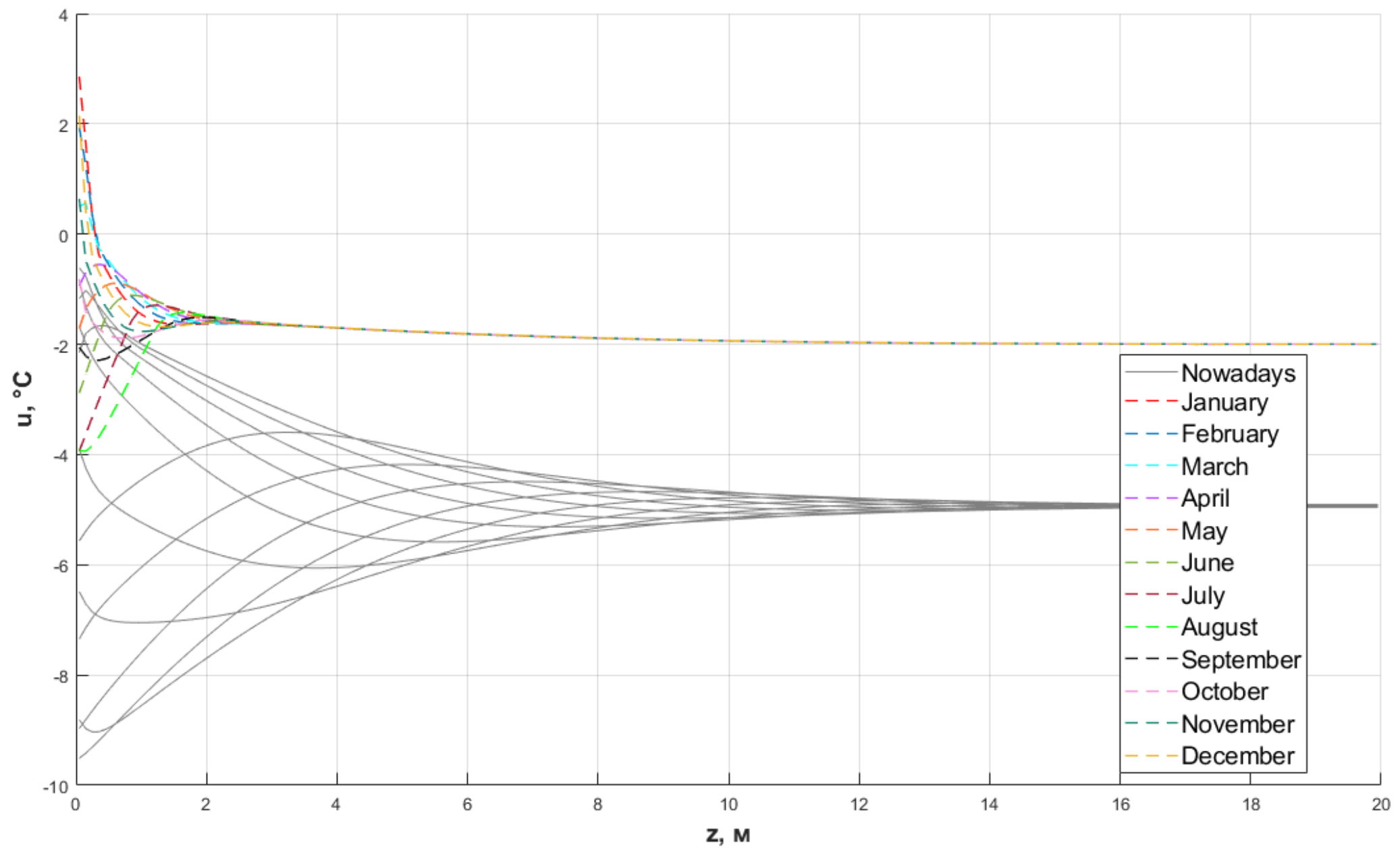

b

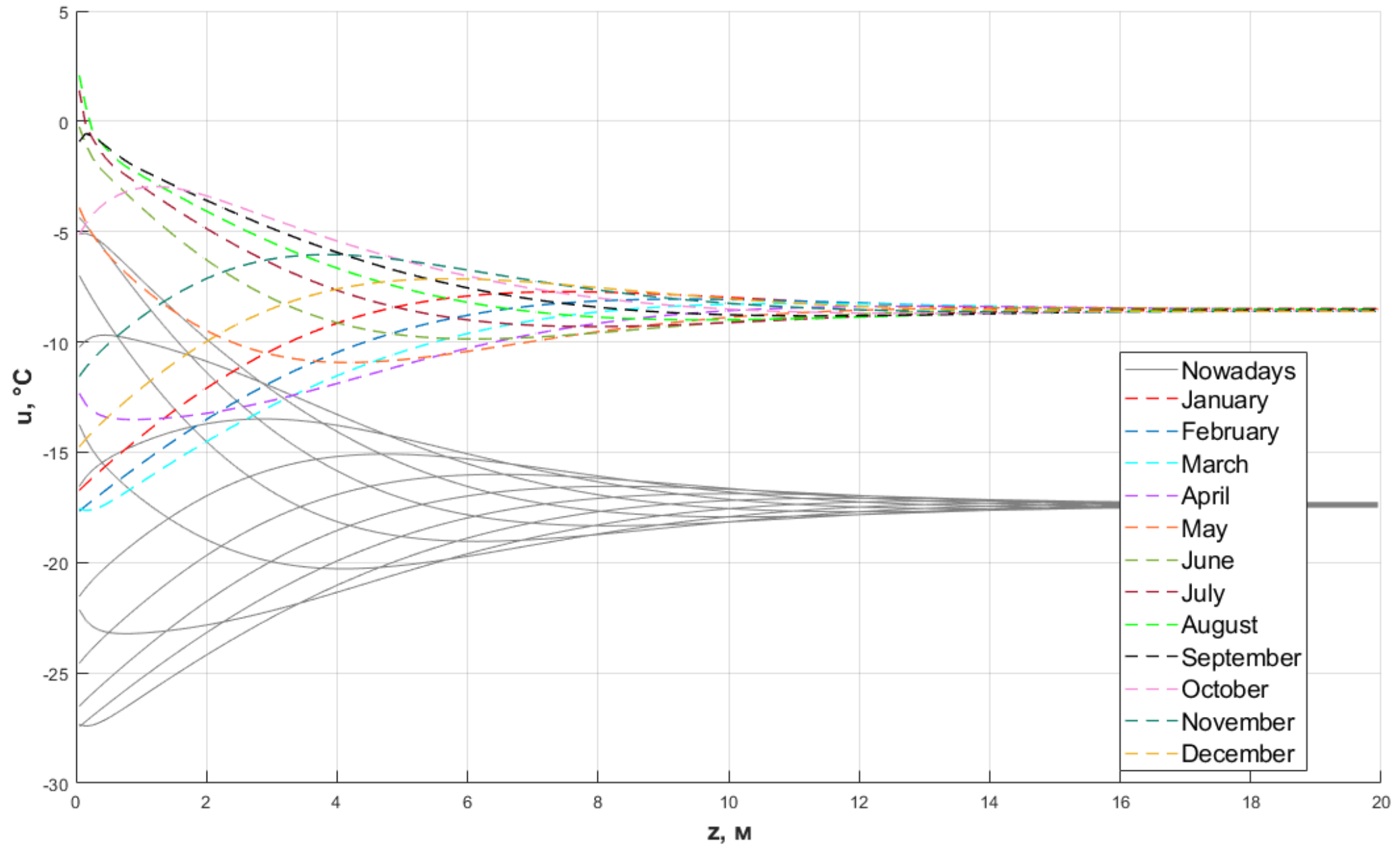

c

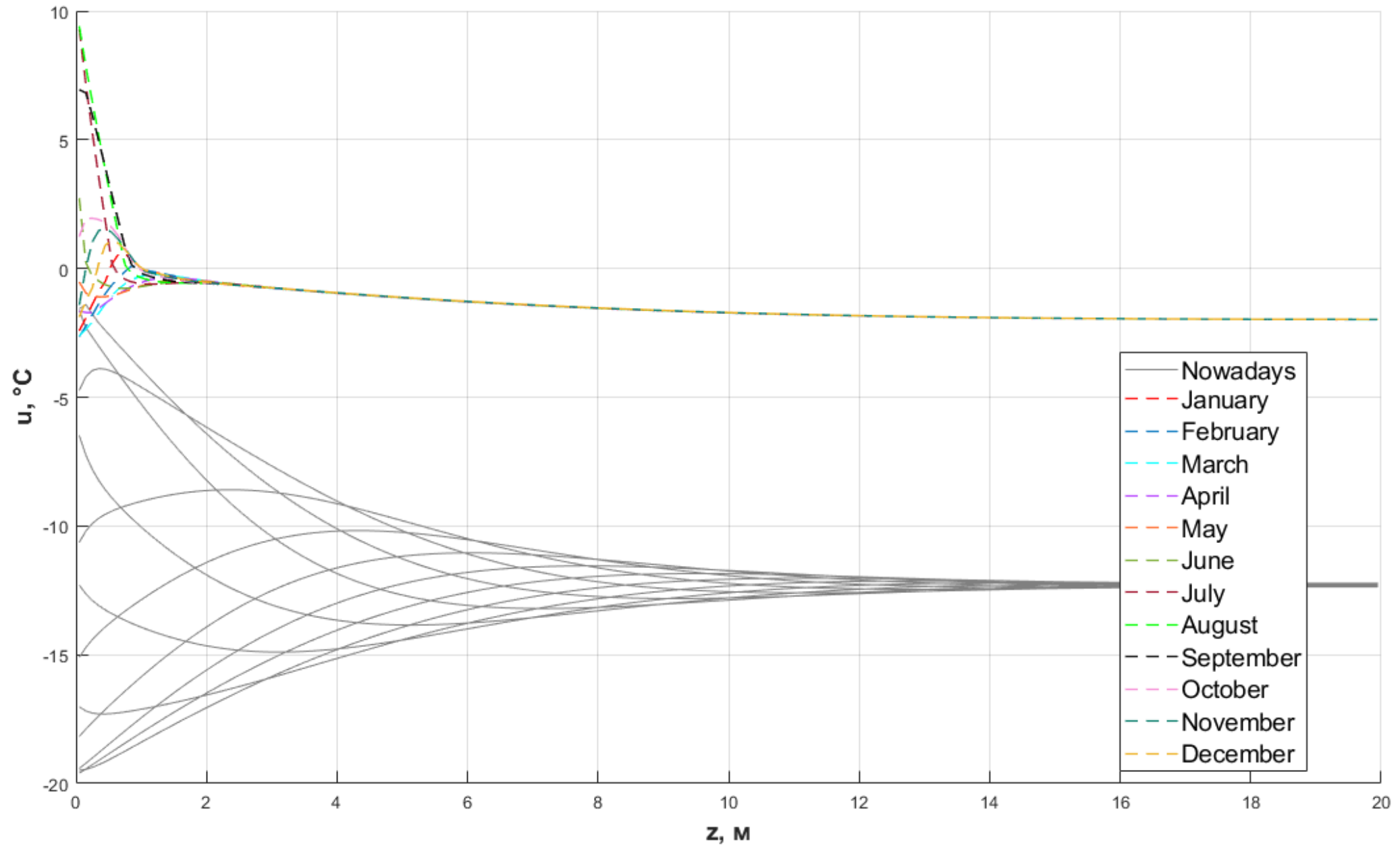

d

Fig. 6. Graphs of non-stationary periodic regimes of ice temperatures nowadays (gray) and predicted for the RCP7 scenario for 2080-2100. (colored): (a) Amundsen-Scott, (b) Esperanza, (c) Cape Morris Jesup, (d) Ernst Krenkel Observatory.

In Antarctica, at Amundsen-Scott South Pole Station (Fig. 6, a) near the surface the ice temperature varies from -54,9 to -24,5°C (was -59,4 to -29°C), at a depth of 4 m it varies from -49.9 to -38.6°C (was -54.5 to -42.9°C). The depth of the zero annual amplitudes actually remained at 16 m. The non-stationary periodic temperature regime shifted by 4,5°C towards heat and reached -45°C.

At Esperanza Base (Fig. 6, b) near the surface the ice temperature varies from -3,9 to 2,9°C (was -9,5 to -0,6°C), at 4 m depth the temperature throughout the year remains equal to -1,7°C (was -6,4 to -3,6°C). In this scenario, an active layer with a depth of 0,3 m was formed. The depth of zero annuals amplitudes decreased to 2 m. The non-stationary periodic temperature regime at a depth of 20 m shifted by 2,9°C towards heat and reached -2°C.

In the Arctic, at Cape Morris Jesup (Fig. 6, c) near the surface the ice temperature varies from -17,7 to 2,1°C (was -27,4 to -4,4°C), at 4 m depth it varies from -11,9 to -5,4°C (was -21,4 to -13,3°C). This scenario produced an active layer of 0,22 m depth. The depth of zero annual amplitudes decreased to approximately 13 m. The non-stationary periodic temperature regime at a depth of 20 m shifted by 8,8°C and reached -8,6°C.

At the Ernst Krenkel Observatory (Fig. 6, d) near the surface the ice temperature varies from -2,7 to 9,4°C (was -19,6 to -1,6°C), at 4 m depth the temperature during the whole year remains equal to -0,96°C (was in the -15,2 to -9°C range). In this scenario, an active layer with a depth of 1 m was formed. The depth of zero annual amplitudes decreased to 1,5 m (was 14 m). The non-stationary periodic temperature regime at a depth of 20 m shifted by 10,3°C and reached a value of - 2°C.

Simulations of the ice column temperature regime for the RCP7 scenario signal severe thawing at the Ernst Krenkel Observatory, Esperanza Base and Cape Morris Jesup stations. Such results demonstrate the urgency of the warming that will occur if current emission trends continue, as it will cause a severe reduction in the ice cover area.

Both scenarios considered indicate the need for a drastic slowdown in climate warming.

## VI. Prediction of ice temperature regime with regard to climate warming constraints

Human activities are estimated to cause global warming of about 1°C above pre-industrial levels, with a probable range of 0,8 to 1,2°C, which characterizes the average warming for a time span of 30 years centered in 2017. This conclusion suggests a likely global warming of 1,5°C by 2050. The challenge for humanity will be to keep it near this mark until 2080-2100, since the already current warming, due to anthropogenic emissions from pre-industrial times to the present, will not stop for hundreds to thousands of years and will continue to cause further long-term changes in the climate system. But it cannot be asserted with certainty that anthropogenic emissions are the only cause of such changes [39].

In 2015, under the UN Framework Convention on Climate Change, the Paris Agreement was adopted, which regulates measures to reduce atmospheric carbon dioxide from 2020. The goal of the agreement is to keep the global average temperature rise "well below" 2°C and "make efforts" to limit the temperature rise to 1,5°C [39].

Continued warming of up to 2°C will increase the risks to natural and human systems further. Risks of drought and precipitation deficits are expected to increase in some regions. In other regions, collectively, heavy precipitation on a global scale will be higher with a global warming of 2°C compared to a warming of 1,5°C, which will increase the land area exposed to flood hazards [1].

Also, by 2100, global mean sea level rise is expected to be about 0,1 m lower with global warming of 1,5°C compared to 2°C. Sea level will continue to rise well beyond 2100, and the magnitude and rate of this rise will depend on the trajectories of future emissions. A slower sea level rise will allow anthropogenic and ecological systems in small islands, low-lying coastal areas, and river deltas to adapt better [1].

It is reasonably certain that impacts on biodiversity and ecosystems, including species decline and extinction, will be less extensive with a global warming of 1,5°C compared to a warming of 2°C [1].

A limited warming of 1.5°C would reduce the increase in ocean temperature, as would the accompanying increase in ocean acidification and decrease in oxygen content, which would reduce risks to marine biodiversity, fisheries and ecosystems and preserve their functions (e.g. ice sheets and coral reefs) [1].

Climate-related risks to health, livelihoods, food security, water security, human security and economic growth are also expected to increase further with a 2°C warming. Most adaptation needs would decrease with a global warming of 1,5°C [1].

For scenarios with no or limited exceedance of 1,5°C, net global anthropogenic CO2 emissions are reduced by about 45% from 2010 levels by 2030, reaching net zero by about 2050. To limit global warming to below 2°C, CO2 emissions are reduced by about 25% by 2030 and reach net zero by about 2070. Emissions of other gases show similar sharp reductions for global warming of 1,5°C and 2°C [1].

Let’s simulate the effect of global warming according to the RCP1.9 scenario on the temperature regime of the ice column. This scenario has similar behavior to the RCP2.6 scenario, that is, the temperature reached through warming by 2050 will remain at this level until 2080-2100. The average global warming in the RCP2.6 scenario is 1.5°C and in RCP1.9 it is 1° relative to current levels (1960-2021) [28].

For this scenario, Table 6 was compiled to show the numerical values of warming for each station by 2080-2100.

Table 6 Numerical warming values for the RCP1.9 scenario.

| Station | Amundsen-Scott | Esperanza | Morris Jesup | E. Krenkel Observatory |
|---|---|---|---|---|
| RCP1.9, °C | 1,25 | 0,75 | 3 | 4 |

On the basis of these values, the graphs for the predictive model of changes in the temperature distribution of the ice column for the RCP1.9 scenario were plotted (Fig. 7).

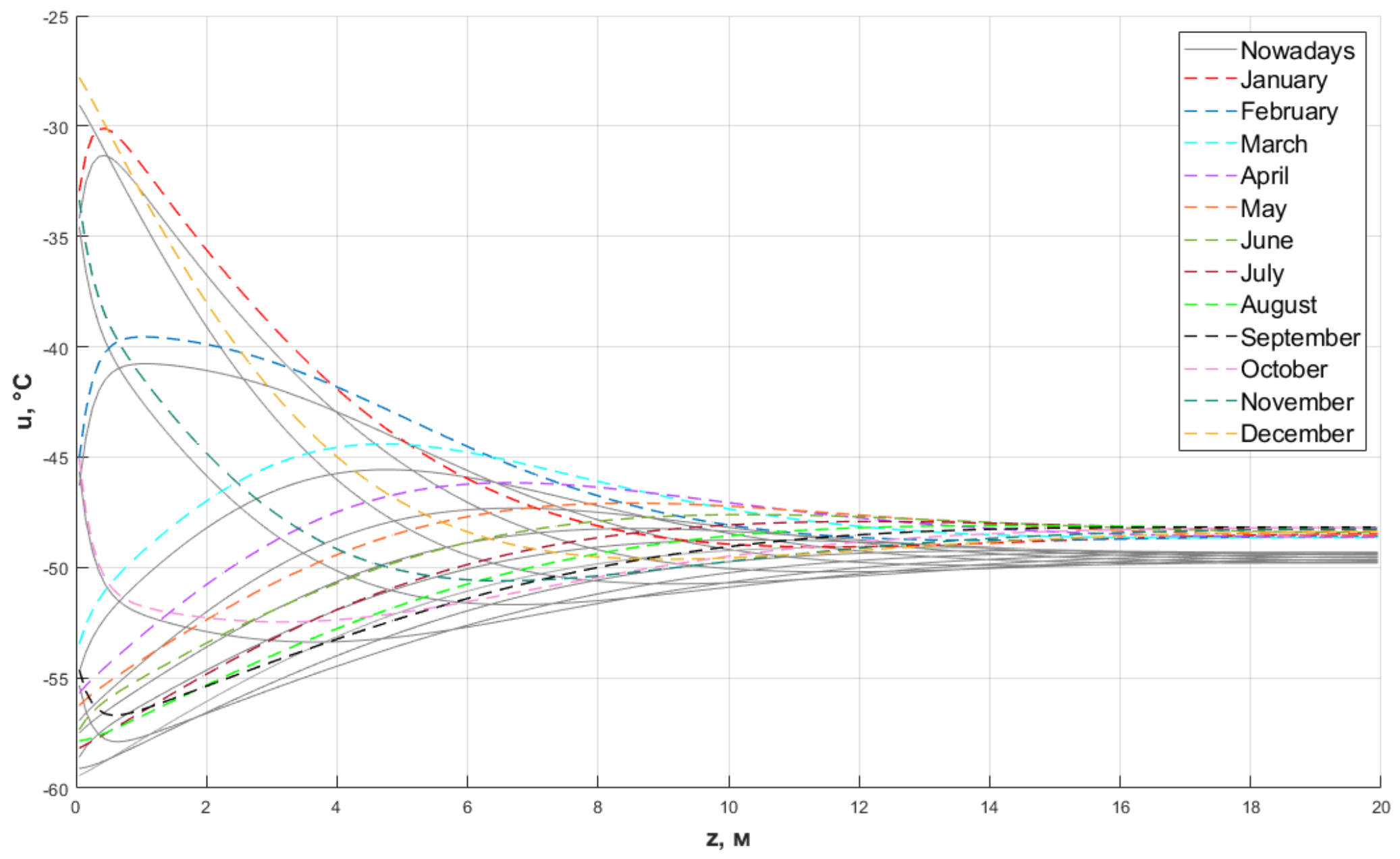

a

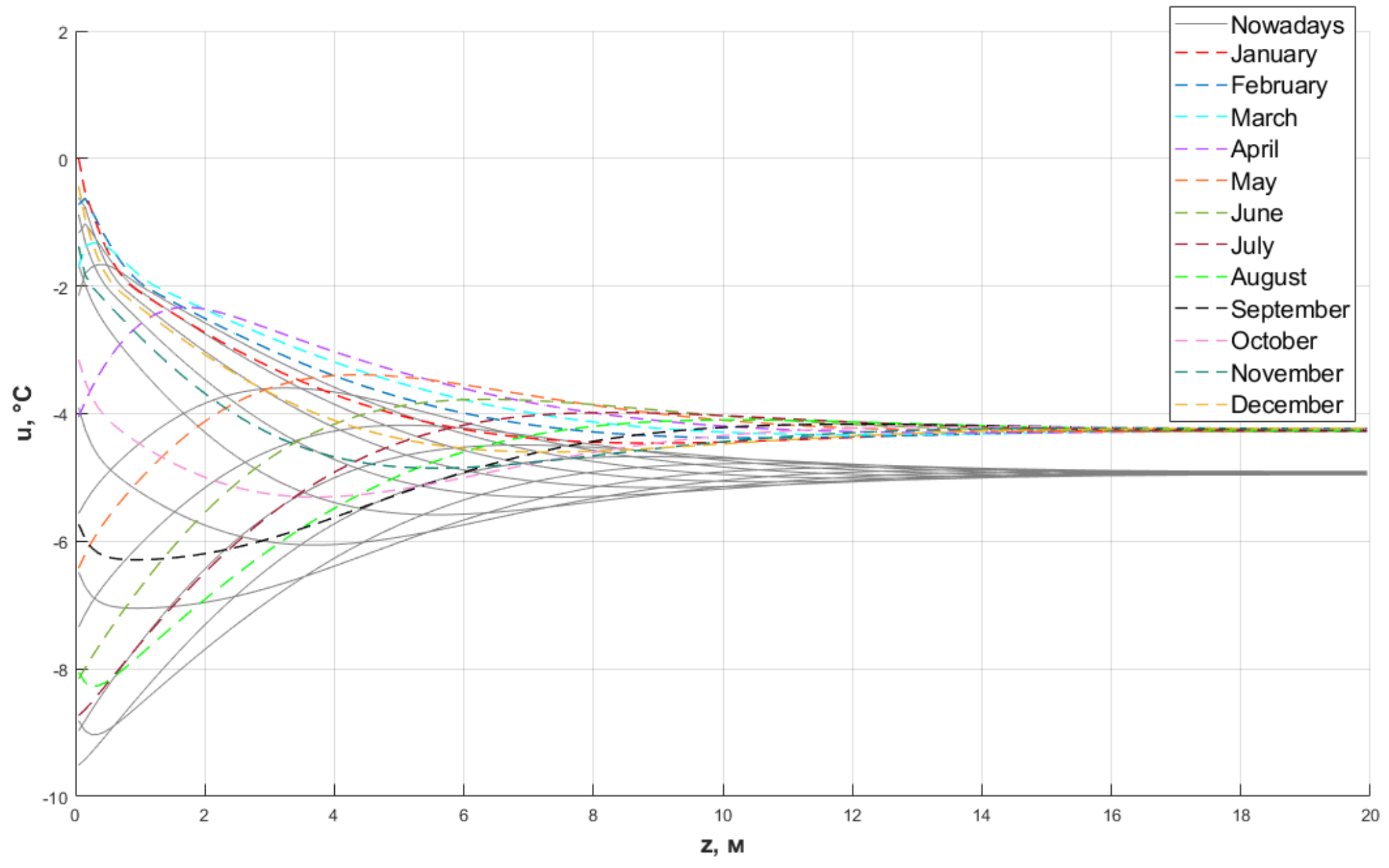

b

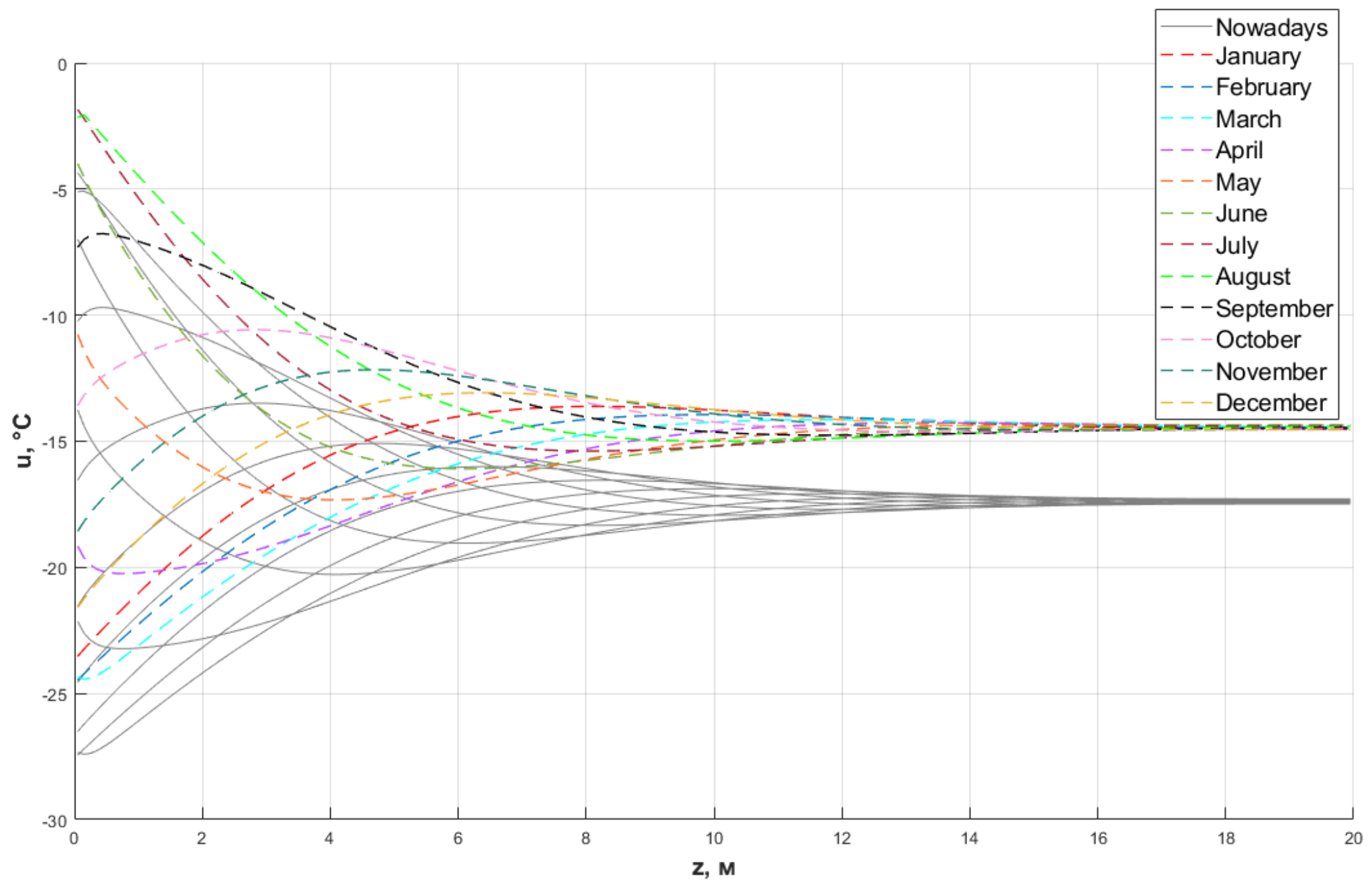

c

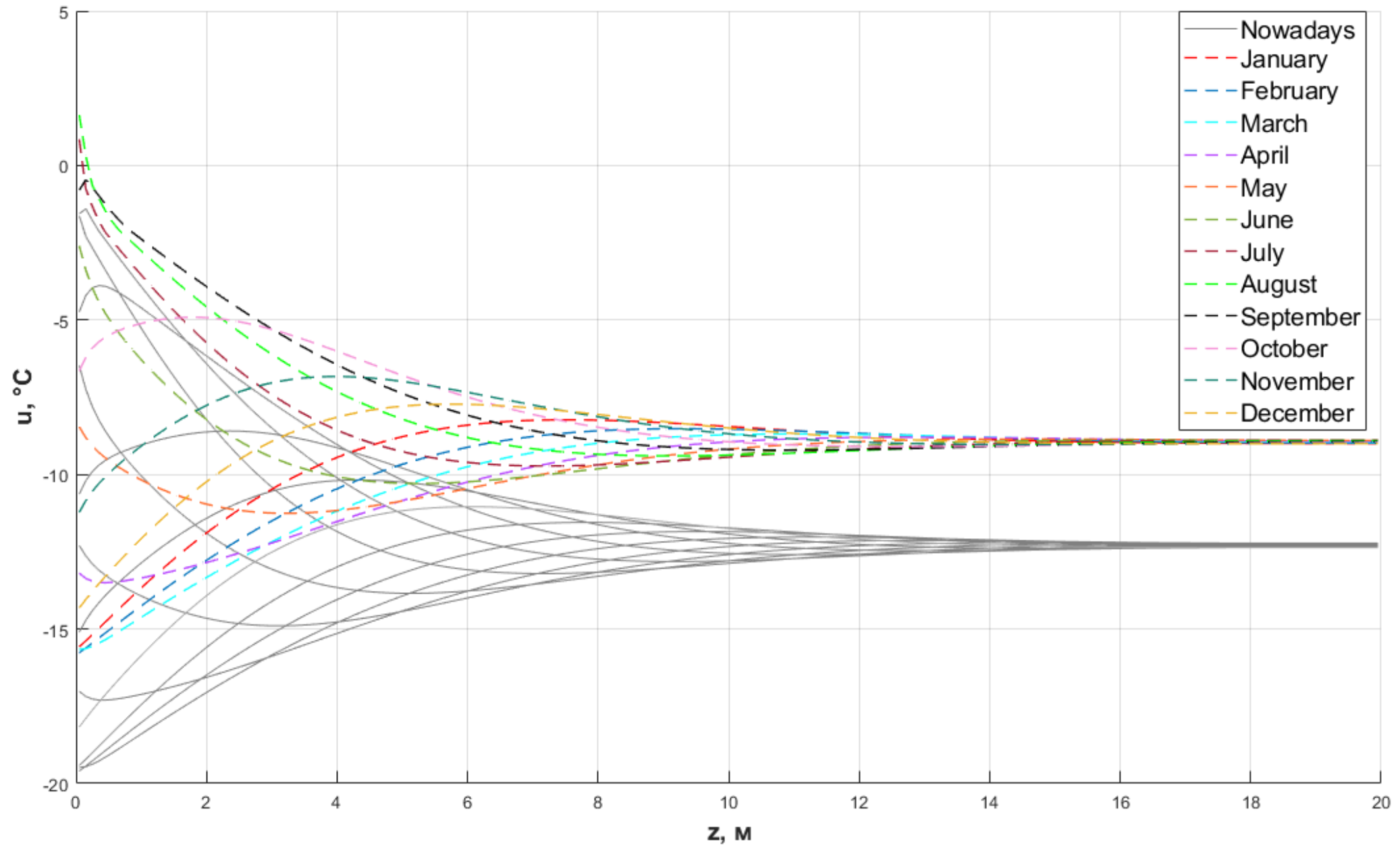

d

Fig. 7. Graphs of non-stationary periodic regimes of ice temperatures nowadays (gray) and predicted for the RCP1.9 scenario for 2080-2100. (colored): (a) Amundsen-Scott, (b) Esperanza, (c) Cape Morris Jesup, (d) Ernst Krenkel Observatory.

In Antarctica, at Amundsen-Scott South Pole Station (Fig. 7, a) near the surface the ice temperature changes in the range from -58,2 to -27,8°C (was -59,4 to -29°C), at a depth of 4 m in the range from -53,3 to -41,8°C (was -54,5 to -42,9°C), the warming atmosphere is reflected on the ice almost linearly due to absence of the snow cover. The depth of the zero annual amplitudes actually remained equal to 16 m. The non-stationary periodic temperature regime shifted by 1,1°C towards warming and reached -48,4°C.

At Esperanza Base (Fig. 7, b) near the surface the ice temperature changes in the range from -8,7 to 0°C (was -9,5 to -0,6°C), at a depth of 4 m in the range from -5,6 to -3°C (was -6,4 to -3,6°C). The depth of zero annual amplitudes remained approximately equal to 14 m. The non-stationary periodic temperature regime has shifted by 0.7°C towards heat and reached the value of -4,2°C.

In the Arctic, at Cape Morris Jesup (Fig. 7, c) near the surface the ice temperature varies from -24,5 to -1,85°C (was -27,4 to -4,4°C), at a depth of 4 m between -18,4 to -10,5°C (was -21,4 to -13,3°C). The depth of zero annual amplitudes remained approximately equal to 14 m. The non-stationary periodic temperature regime has shifted by 2,9°C and reached the value of -14,5°C.

At the Ernst Krenkel Observatory (Fig. 7, d) near the surface the ice temperature changed in the range from -15,8 to 1,6°C (was -19,6 to -1,6°C), at a depth of 4 m in the range from -11,5 to -6°C (was in the range from -15,2 to -9°C). In this scenario, an active layer appears with a depth of 0,18 m, i.e. ice thawing is observed. The depth of the zero annual amplitudes decreased to 13 m (it was 14 m). The non-stationary periodic temperature regime shifted by 3,3°C and reached -9°C.

The simulations performed for the years 2080-2100 of the RCP1.9 scenario demonstrated the appearance of an active layer only at the Ernst Krenkel Observatory in the Arctic. In Antarctica at the Esperanza Base, warming led to the ice reaching 0°C in January, indicating that the thawing process is just beginning.

# VII. Conclusion

In this work, a model of the ice temperature regime at two stations located near the South Pole and two stations near the North Pole was constructed. The temperature plots for each month were constructed and the depths of active layer (freezing/thawing) and the depths of zero annual amplitudes were determined. A model of the ice temperature regime for our time, as well as three prognostic models for global warming scenarios for the years 2080-2100 were compiled.

Calculation results for the RCP2.6 and RCP7 scenarios show significant changes in the temperature regime of the pure freshwater ice column. However, the calculations obtained for RCP1.9 scenarios clearly demonstrate the positive effect of restraining warming at 1,5°C from pre-industrial levels. For this scenario, one can expect a sufficient slowdown in the reduction of the ice cover area and the melting of ice sheets, as well as a reduction in numerous risks to people and various ecosystems of our planet.

Anatoliy Fedotov, Department of Mathematics, Bauman Moscow State Technical University (5/1 2-nd Baumanskaya St., Moscow 105005, Russia)

*Email address*: fedotov_a_a@bmstu.ru

Vladimir Kaniber, Department of Mathematics, Bauman Moscow State Technical University (5/1 2-nd Baumanskaya St., Moscow 105005, Russia)

*Email address*: kanibervv@student.bmstu.ru

Pavel Khrapov, Department of Mathematics, Bauman Moscow State Technical University (5/1 2-nd Baumanskaya St., Moscow 105005, Russia)

*Email address*: khrapov@bmstu.ru